\documentclass[12pt]{article}
\usepackage{authblk}
\usepackage[hidelinks]{hyperref}
\usepackage{comment}
\usepackage[T1]{fontenc}
\usepackage[utf8]{inputenc}
\usepackage{mathpazo}

\usepackage[a4paper, margin=2cm]{geometry}
\usepackage{graphicx}
\usepackage{enumitem}
\usepackage{textcomp}
\usepackage{varwidth}
\usepackage{setspace}
\usepackage[font=normalsize]{caption}
\usepackage{multicol, multirow}
\usepackage[justification=centering]{caption}
\usepackage{color, colortbl, soul}
\usepackage{amsmath}
\usepackage{gensymb}
\usepackage{tabu}
\usepackage{booktabs}
\usepackage{cancel}
\usepackage{cleveref}
\usepackage{listings}
\usepackage{pgfplots}
\usepackage{xcolor}
\usepackage{siunitx}
\usepackage{multicol} 
\usepackage{tabularx}

\usepackage[title]{appendix}

\makeatletter
\renewcommand\appendix{\par
\setcounter{section}{0}%
\setcounter{subsection}{0}%
\setcounter{equation}{0}%
\setcounter{table}{0}
\setcounter{figure}{0}
\gdef\theequation{\@Alph\c@section.\arabic{equation}}%
\gdef\thefigure{\@Alph\c@section.\arabic{figure}}%
\gdef\thetable{\@Alph\c@section.\arabic{table}}%
\gdef\thesection{\appendixname\@Alph\c@section}%
\@addtoreset{equation}{section}%
\@addtoreset{table}{section}
\@addtoreset{figure}{section}
}
\makeatother

\newcommand{\overbar}[1]{\mkern 1.5mu\overline{\mkern-1.5mu#1\mkern-1.5mu}\mkern 1.5mu}

\usepackage[colorinlistoftodos]{todonotes}

\usepackage[modulo]{lineno}

\newcommand{\fref}[1]{Figure~\ref{#1}}

\newcommand{\eref}[1]{Equation~\ref{#1}}
\newcommand{\sref}[1]{Section~\ref{#1}}

\newcommand{\jwunit}[1]{\ensuremath{\, \si{#1}}}
\newcommand{\micron}{\ensuremath{\, \si{\micro \metre}}}

\hypersetup{
colorlinks = true,
linkcolor = black, 
citecolor = black, 
urlcolor = blue, 
filecolor = black 
}

\usepackage{natbib}
\setcitestyle{authoryear,open={(},close={)}}

\usepackage{todonotes}
\usepackage{soul}

\usepackage{abstract}

\newcommand{\x}{\boldsymbol{x}}

\newcommand{\Nsl}{N_{SL}}

\begin{document}

\doublespacing

\title{Validated respiratory drug deposition predictions from 2D and 3D medical images with statistical shape models and convolutional neural networks}
\author[1]{Josh Williams}
\author[1]{Haavard Ahlqvist}
\author[1]{Alexander Cunningham}
\author[2]{Andrew Kirby}
\author[3]{Ira Katz}
\author[4,5]{John Fleming}
\author[4,6]{Joy Conway}
\author[7]{Steve Cunningham}
\author[1*]{Ali Ozel}
\author[1*]{Uwe Wolfram}

\affil[1]{School of Engineering and Physical Sciences, Heriot-Watt University, Edinburgh, UK}
\affil[2]{Royal Hospital for Children and Young People, NHS Lothian, Edinburgh, UK}
\affil[3]{Consultant, Meudon, France}
\affil[4]{National Institute of Health Research Biomedical Research Centre in Respiratory Disease, Southampton, UK}
\affil[5]{Department of Medical Physics and Bioengineering, University Hospital Southampton NHS Foundation Trust, Southampton, UK}
\affil[6]{Respiratory Sciences, Centre for Health and Life Sciences, Brunel University, London, UK}
\affil[7]{Centre for Inflammation Research, University of Edinburgh, Edinburgh, UK}
\maketitle
\thanks{* UW and AO share last authorship. \\
Address correspondence to u.wolfram@hw.ac.uk or a.ozel@hw.ac.uk}
\newpage

\begin{abstract}

For the one billion sufferers of respiratory disease, managing their disease with inhalers crucially influences their quality of life.
Generic treatment plans could be improved with the aid of computational models that account for patient-specific features such as breathing pattern, lung pathology and morphology.
Therefore, we aim to develop and validate an automated computational framework for patient-specific deposition modelling. 
To that end, an image processing approach is proposed that could produce 3D patient respiratory geometries from 2D chest X-rays and 3D CT images.
We evaluated the airway and lung morphology produced by our image processing framework, and assessed deposition compared to \textit{in vivo} data.
The 2D-to-3D image processing reproduces airway diameter to 9\% median error compared to ground truth segmentations, but is sensitive to outliers of up to 33\% due to lung outline noise.
Predicted regional deposition gave 5\% median error compared to \textit{in vivo} measurements.
The proposed framework is capable of providing patient-specific deposition measurements for varying treatments, to determine which treatment would best satisfy the needs imposed by each patient (such as disease and lung/airway morphology). 
Integration of patient-specific modelling into clinical practice as an additional decision-making tool could optimise treatment plans and lower the burden of respiratory diseases.


\end{abstract}

\newpage
\section{Introduction}
\label{sec:intro}    

Respiratory diseases affect over one billion people worldwide \citep{globalAsthma2019}. 
Diseases such as asthma, cystic fibrosis and COPD are typically treated with inhaled therapeutics which are delivered to the affected tissue to prevent symptoms such as breathlessness from arising, or relieve them when they do arrive. 
Many patients' symptoms are worsened by a combination of poor inhaler technique and a lack of adherence to their treatment plan \citep{van2018personalising}. 
Electronic health (e-health) measures such as smart inhalers \citep{van2018personalising} are being introduced to improve adherence and quality of care \citep{honkoop2022current}.
Smart inhalers can use sensors to track how often patients use inhalers and also provide estimates on technique or inspiratory flowrate based on acoustic sensors \citep{van2018personalising, lim2019robust}.
The value of e-health monitoring for patients could be increased significantly by linking measurements such as inhalation flowrate with patient-specific physical models that accurately predict how the inhaled drug has deposited in patient lungs.

Patient-specific modelling of drug deposition can be done with computational particle-fluid dynamics (CPFD) which solves the governing equations of air and drug particle transport during inhalation \citep{nowak2003computational, koullapis16particle, williams2022effect}.
Patient-specific CPFD modelling requires 3D representations of the patient's respiratory system, usually obtained from clinical computed tomography (CT) scans \citep{kleinstreuer2010review}.
Typically the airways are segmented using manual approaches, which can take several hours \citep{sonka1996rule}, and is therefore unsuitable for usage in future e-health monitoring frameworks. 
\citet{tschirren2009airway} showed that manually correcting airway segmentations obtained from semi-automatic methods can also take several hours. 
Such a long and labour-intensive process is a substantial bottleneck in applying patient-specific deposition models to a large number of respiratory patients. 
Fully-automatic segmentation methods have recently become available, which are promising for rapid segmentation of lungs and airways from volumetric CT using convolutional neural networks (CNNs) \citep{garcia2021automatic, hofmanninger2020automatic, yun2019improvement, tang2019automatic}. 
CNNs apply varying levels of convolution, where the convolution kernels are learned by evaluating the difference between the CNN output compared to a `ground truth' human-verified result \citep{lecun1995convolutional}. 
CNNs have been widely used in medical image processing and are used here to segment airways from 3D images such as clinical CT. 
However, CT images are more costly to acquire than chest X-rays and are therefore not optimal for imaging a large number of respiratory sufferers.
Additionally, using CT scans is not encouraged in children for routine assessments due to harmful radiation. 
Therefore, an additional image-processing approach that can extract 3D respiratory systems from sparser imaging such as chest X-rays is an unmet need.

To perform deposition simulations based on patient chest X-rays, we require a novel image-processing approach to reconstruct 3D respiratory geometries from 2D chest X-rays.
Reconstruction of 3D skeletal anatomies including femora, vertebra or knee joints from planar and bi-planar X-ray images has been performed using statistical shape and appearance models (SSAMs) \citep{albrecht2013posterior, baka20112D3D, vaananen2015generation, zheng2011scaled, youn2017iterative, dworzak20103d}. 
Statistical shape models (SSMs) take a set of corresponding geometries (usually represented with landmarks that represent key features), and quantifies how shape features change across a population (based on eigenvalues and eigenvectors of the covariance matrix) \citep{cootes1995active}. 
A SSAM performs the same operation, including an additional feature such as gray value at each landmark.
In this approach, patient lungs may be extracted from a chest X-ray by combining (i) available information such as lung outline and gray-value distribution, as well as (ii) a SSAM that describes how lung shapes and projected gray-value varies from a database of known (3D) lung shapes \citep{cootes1995active, cootes2001active, heimann2009statistical}. 
Respiratory statistical shape models (SSMs) have previously been used for diagnostic purposes, as \citet{irving20132Dxray} developed an approach to extract 3D airways from a single chest X-ray and diagnose tuberculosis in children \citep{irving2011segmentation, irving20132Dxray}. 
\citet{osanlouy2020lung} used a SSM of the lungs to quantify differences in lung morphology in smokers and non-smokers. 
These examples of SSM and SSAM applications in skeletal and respiratory imaging show that SSAMs are a suitable approach for extracting full respiratory systems from planar or bi-planar chest X-rays. 
However, a key issue in this approach is that only the first bifurcation-level (or `generation') of the airway tree is visible on a chest X-ray \citep{irving20132Dxray}, as the low density lung structures are overlapped by denser structures such as the spine, rib cage and heart \citep{gozes2018lung, yang2017cascade, liang2020bone}. 
Similarly, only around 5-10 generations (out of 23) are visible on high resolution CT scans \citep{kleinstreuer2010review, bordas2015}. 
Thus, even with high resolution imaging, the deep lung airways are left unaccounted for and must be included for patient-specific deposition predictions.

The limitations of imaging resolution have been overcome in several ways. The remaining levels of the conducting airway tree can be generated (approximately up to generation 16) using statistical models based on information from the visible airways \citep{tawhai2004, bordas2015, montesantos2016creation}. 
The remaining portion of the airway tree (bifurcations 17 to 23) are the respiratory `acinar' region, which is a very small (characteristic length is $\mathcal{O}(100 \jwunit{\micro m})$) and intricate structure that is impossible to image \textit{in vivo} \citep{weibel1988morphometry, rausch2011local, dong2022recent}. 
Acinar flows have been studied with CPFD \citep{sznitman2013respiratory, dong2022recent, khajeh2015deposition, hofemeier2016role}, but it is highly computationally expensive to directly resolve the full conducting airways and deep lung in one simulation \citep{koul18efficomp, koullapis2019multiscale, koullapis2020deeplung}. 
Typically, the deep lung mechanics are treated with reduced-dimension models such as 1D airway networks or 0D lumped parameter models which are coupled to 3D CPFD simulations at the 3D airway outlet \citep{oakes2015distribution, oakes2018airflow, roth2017computational, comer10structuredtree, ismail2014stable, pozin2017tree, kuprat2020efficient, poorbahrami2019regional, feng2021tutorial}. 
To model the full respiratory system, we therefore couple 3D CPFD domains produced from our image-processing framework with a 0D lumped parameter model to validate the entire image-processing and simulation framework against experimental deposition data \citep{conway2012controlled, fleming2015controlled}.

Therefore, we aimed to (i) develop a computational framework to reconstruct 3D airways and lungs from 2D image(s); (ii) develop a segmentation tool for reconstructing airways and lungs from clinical CT data; (iii) evaluate the ability of the developed image-processing tools to reproduce morphological properties and deposition of the ground truth airways, and (iv) validate deposition predictions against experimental data. Our developed computational framework is made publicly available online (\url{https://github.com/jvwilliams23/respiratory2Dto3Dpaper}).

\section{Methods and materials} \label{sec:methods}

\subsection{Patient image data and ground truth generation} \label{sec:dataset}
We used 51 clinical CT scans from the LUNA16 challenge \citep{setio2017luna16}, which is part of The Cancer Imaging Archive (TCIA) \citep{clark2013tcia}. 
The scans were acquired for lung cancer nodule detection which does not depend on inhalation state, therefore no information on breath-hold procedures was reported by \citet{setio2017luna16}.
The scans used were chosen randomly by \citet{tang2019automatic}, and they provided radiologist-verified lung lobe segmentations online. 
We created airway segmentations for the same scans using a semi-automatic region-growing approach based on \citet{nardelli2015}. These segmentations have been independently verified by a radiologist with 9 years experience. 
We excluded 13 samples from our SSAM dataset, as the segmentations only included the trachea and first bifurcation.

To validate our modelling approach, we compared to experimental data of radiopharmaceutical aerosol deposition from \citet{conway2012controlled} and \citet{fleming2015controlled}, which we refer to throughout as the `Southampton/Air Liquide dataset'. 
\citet{conway2012controlled} and \citet{fleming2015controlled} performed experiments with inhaled aerosol in six healthy and six asthmatic subjects.
Patients sat in an erect (sitting) position while aerosol was delivered through an AKITA nebuliser.
The AKITA nebuliser is breath-actuated to release drug only during inhalation, and a controlled (constant) inhalation profile to optimise deposition \citep{munro2020using}.
Deposition was imaged with single photon emission computed tomography (SPECT) and combined with low resolution X-ray CT captured in the same machine (SPECT-CT), which relates deposition images to anatomical structure of the lungs \citep{fleming2011use}.
To relate deposition measurements to airway morphology, high resolution CT was acquired for each patient on a different machine on a different day \citep{conway2012controlled, montesantos2013airway}. 
Full information on scanning equipment and protocol are provided elsewhere \citep{conway2012controlled, montesantos2013airway}, and only key points related to our study are discussed here.
High resolution CT images were acquired in supine position, which is known to decrease lung volume compared to erect position \citep{ibanez1982normal}.
Additionally, \citet{conway2012controlled} discussed imaging issues in the upper airway. 
In half of the images acquired, the epiglottis was completely closed, meaning there was no path for air to pass from the mouth to the lung airways.
This may have occurred since the images were taken during exhalation, where the glottis area is at its narrowest \citep{brancatisano1983vocalcords, scheinherr2015realistic}.
Furthermore, motion artefacts may have blurred the already narrow glottis to appear fully closed.
To allow for air and drug particles to pass from the mouth to the lung, we manually cleaned up segmentations in these problematic cases (Supplementary Material Section 1).

\subsection{Shape reconstruction from a chest X-ray image}

To reconstruct 3D lung and airway geometries from an X-ray image, we created a statistical shape and appearance model (SSAM) based on ground truth airway and lung lobe segmentations.
The appearance is based on digitally reconstructed radiographs (DRRs) derived from the patient CT data (described below).
The DRRs were also used for testing how well the SSAM fits to an X-ray.
Below we describe the process to create a SSAM from patient CT data and segmentations described above (\sref{sec:dataset}).

\subsubsection{Pre-processing and landmarking} \label{sec:preproc}
To establish correspondence amongst the lobes, we make use of the \textit{`Growing and Adaptive Meshes'} (GAMEs) algorithm of \citet{ferrarini2007games}. This algorithm determines the set of landmarks needed to best describe the surface mesh by clustering points based on their proximity and covariance. A set of landmarks are grown for a base mesh that acts as a population mean~\citep{marsland2002self}. 
The base landmarks are then iteratively adapted to fit the rest of the population, such that adapted points correspond to the same feature \citep{kohonen1990self, kohonen2013essentials}. 
Outlier points that had not correctly fitted to a new shape were detected as points with a Mahalanobis distance of greater than 0.15, which we determined empirically by analysing statistics of poorly fitted cases. 
Once all points were below this threshold, the landmark adaptation was considered converged. If convergence was not satisfied and the maximum distance has settled at a steady-state (change in maximum Mahalanobis distance less than 0.05 after five full iterations through all points in ground truth), then the landmarking was considered failed. If landmarking failed, another cloud of landmark points was used as a base to adapt from.

As we cannot directly see the lobar fissures on a chest X-ray, we removed landmarks on the lobar fissures from the landmark list. Landmarks were classified as being on a lobar fissure if the (dimensionless) Mahalanobis distance between a landmark in one lobe and a landmark in the adjacent lobe was less than 0.2. 
This allowed the SSAM to capture more variance on the outer curvature of the lungs instead of the highly variable lobar fissures, which improves adaptation to lung outlines on X-ray images.
Landmarks for all lobes were combined into one dataset to have a SSAM for the entire lungs (instead of individual lobes), which allows us to capture inter-organ relations \citep{cerrolaza2019multiorgan} such as distance between both lungs, orientation and non-overlapping structure of lobes.
This gave us 2966 landmarks in total for the lungs.

To landmark the airways we first skeletonised the segmentation in scikit-image \citep{lee1994building}. 
The skeletonised labelmap was then converted to spatial coordinates with the Skan package \citep{skan}. 
A graph was created from coordinates output from Skan using Networkx \citep{networkx}. 
Branches with length less than $5\jwunit{mm}$ were removed as recommended by \citet{tschirren2005matching}. 
We included the trachea, main and lobar bronchi in the skeleton landmarks, as airways below this will not be visible on the X-ray image.

At each node on the graph we computed the diameter based on the cross sectional area of the 3D segmentation at that location. This was used to get landmarks on the surface by creating a circle centered on and perpendicular to the medial axis. 
We computed a vector from the circle center to each point along its circumference. We found where this vector intersected the surface mesh of the segmentation and set a landmark here. This allowed us to landmark irregular cross-sections. The airway landmarks were combined with lung landmarks, giving us 3486 landmarks in total.

We created DRRs for all patients in the LUNA16 dataset which we used to train the SSAM and assessing performance. This was done by integrating the Hounsfield units for each voxel in the anterior-posterior direction of the scan to create a frontal chest X-ray. We also created additional DRRs in the sagittal plane to test the influence of including an additional projection on the morphologies predicted by the SSAM. The gray-value at the landmark location could then be used in the appearance part of the SSAM. The DRRs represent patient X-rays at the same time point that the CT data used to extract landmarks was taken. The DRRs were also used for assessing our SSAM performance, as we can compare the 3D reconstructions of the lungs and airways from the SSAM to CT at the same instant. By comparing lungs and airways at the same time instant, our results are not influenced by changes in lung or airway shape during breathing.

Similar to \citet{baka20112D3D}, we computed a Canny edge map of the chest X-ray to use for guiding our SSAM fitting.
To reduce noise in the edge map, we first downsampled the X-ray to be four times coarser and applied global histogram equalization. 
We then applied a Canny edge map \citep{canny1986computational}, where the Gaussian kernel has a width of two pixels.
We found this combination of pre-processing steps to produce an edge map with the lowest amount of noise from overlapping structures such as the rib cage.

\subsubsection{Statistical shape and appearance modelling} \label{sec:methods-ssam}
\begin{figure}
    \centering
    \includegraphics[width=\linewidth]{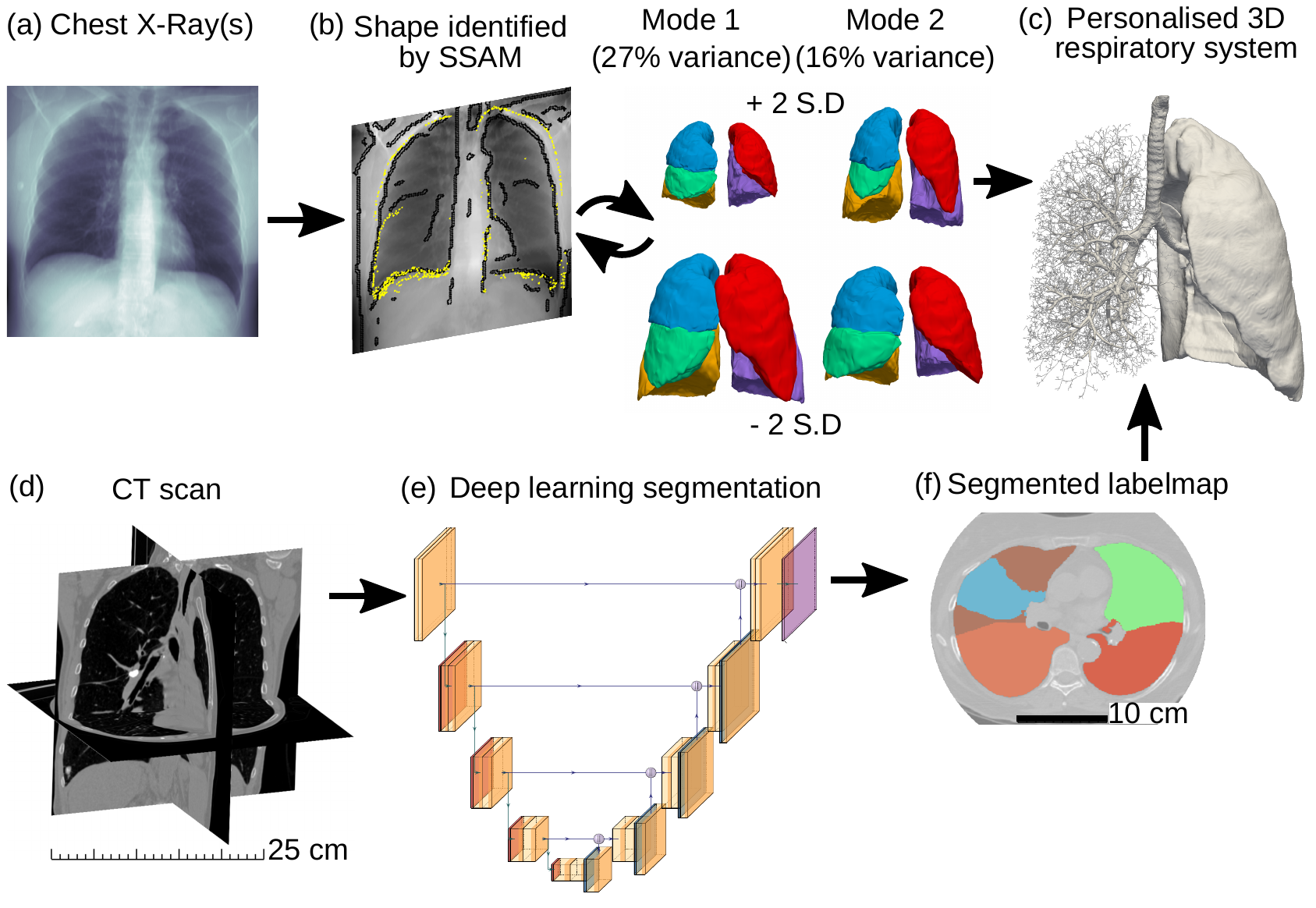}
    \caption{Schematic overview of image-processing framework. Chest X-ray image(s) are provided as an input (a), where the shape is identified by our SSAM by iteratively adapting lung outline and gray-value based on a database of known lung shapes (b-c). The SSAM reconstruction can then be used to generate a surface mesh of lungs with a full airway tree (c). Alternatively, a 3D CT scan may be provided as an input (d) to our trained CNN architecture (e). The CNN returns a segmented labelmap for lungs and airways (f), which can also be used to generate the full respiratory domain (c).}
    \label{fig:framework}
\end{figure}

Once a set of landmarks and DRRs were extracted from the LUNA16 dataset (\sref{sec:preproc}), all sets of landmarks were aligned to zero mean and isotropic scaling was applied such that the coordinates had zero mean and unit standard deviation.
Landmark gray values were also normalised to zero mean and unit standard deviation per sample. 
We then create a SSAM to describe and parameterise the covariance of the lung and airway shape-appearance data in the LUNA16 dataset.
The SSAM parameters are then iteratively adjusted based on how well the lung and airway shape and appearance agrees with the X-ray image (\fref{fig:framework}a-b).

\begin{figure}
    \centering
    \begin{tabular}{ccc}
        (a) & (b) & (c) \\
        \includegraphics{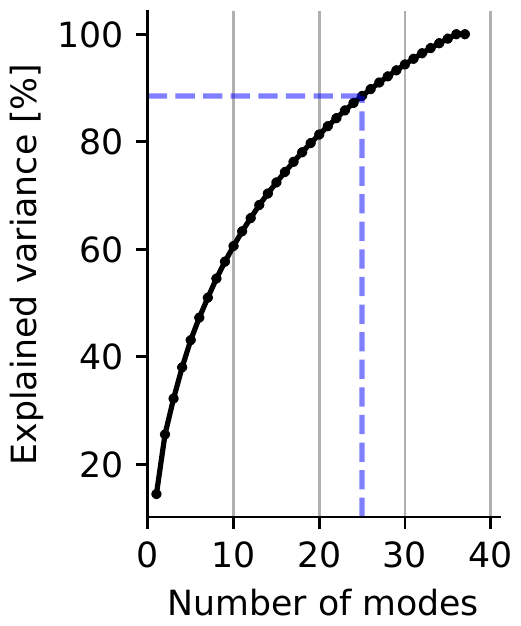} &
        \includegraphics{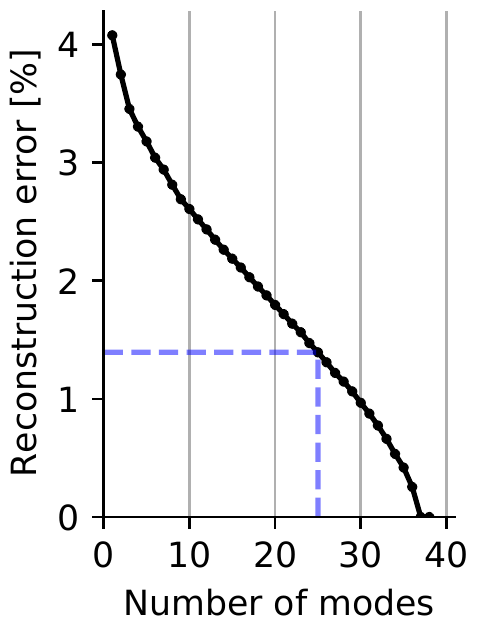} &
        \includegraphics{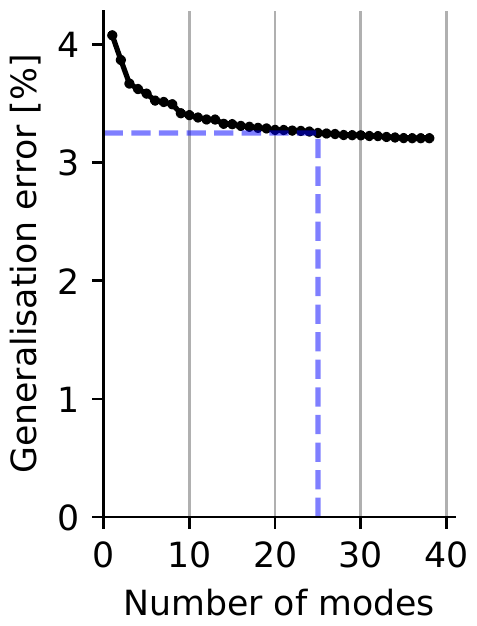}
    \end{tabular}
    \caption{Assessment of sensitivity of SSAM metrics to number of modes included in the model. Panels show (a) explained variance based on number of modes, (b) reconstrucion error for shapes in training set, (c) generalisation error for unseen shapes. When the model describes 90\% of the population variance (indicated by blue dashed line), the generalisation error in adapting the model to unseen shapes is saturated.}
    \label{fig:ssamMetrics}
\end{figure}

The SSAM uses principal component analysis (PCA) to create a model of the variation in the shape and the gray-values of the corresponding chest X-ray. The shape and gray-value, $\boldsymbol{x}$ can then be described by the linear function 
\begin{equation}
    \boldsymbol{x} \approx \boldsymbol{\overbar{x}} 
    + \sum_{m=1}^{N_m} \boldsymbol{\Phi}_{m} \cdot \left( \boldsymbol{b}_m \sqrt{\boldsymbol{\sigma}^2_m} \right),
    \label{eq:ssam}
\end{equation}
where $\boldsymbol{\overbar{x}}$ is a vector of mean shape and gray-value. It has a size of $(3+N_{XR})N_L$, where $N_L$ is number of landmarks, corresponding to a Cartesian coordinate for each landmark and $N_{XR}$ is the number of X-rays per patient, which each contains a set of gray-values for each landmark. 
$\boldsymbol{\Phi}$ is a $(3+N_{XR}) N_L \times N_m$ matrix, giving the main modes of variation of the training set, where $N_m$ is the number of modes. 
$\boldsymbol{\sigma}^2_m$ is the variance described by mode $m$.
$\boldsymbol{b}_m$ is the shape parameter for mode $m$, where \fref{fig:framework}(b) shows the effect of varying the weighting of $\boldsymbol{b}$ for the first two modes.
By using PCA, the complexity of the model can be reduced by including only the number of modes, $N_m$, that are required to capture a specified amount of variance in the modelled shape. 
In this study, 90\% of the population variance is included in the model ($N_m=25$, \fref{fig:ssamMetrics}a).
The SSAM was created using our Python library `\texttt{pyssam}' for statistical shape and appearance modelling \citep{williams2023pyssam}.

The accuracy obtained by the SSAM with $N_m$ modes can be assessed in terms of its ability to reconstruct a shape in the training dataset (reconstruction error). 
More importantly, the SSAM can be assessed by its ability to generalise and reconstruct an unseen shape (not in the training set), which is known as generalisation error.
When the shape model contains 100\% of the training population variance (\fref{fig:ssamMetrics}a), the reconstruction error should tend to zero as all information required to reconstruct the shape is contained within the model (\fref{fig:ssamMetrics}b).
We assessed reconstruction and generalisation error with leave-one-out testing, which was calculated over the entire training set and repeated 30 times.
The mean absolute error is presented, which is expressed as a percentage of the lung bounding-box size.
We observed that $N_m=25$ is a suitable choice to reduce the SSAM dimensionality, as the generalisation error did not increase further by including additional modes (\fref{fig:ssamMetrics}c).

The shape parameters, $\boldsymbol{b}$, can then be optimised with respect to the fit of the SSAM outline ($\mathcal{L}_{fit}$) and the similarity to the training set ($\mathcal{L}_{prior}$). The appearance model optimises the shape by minimising the gray-value loss ($\mathcal{L}_{g}$). Gray-value loss is calculated as the difference between the image gray-value at the landmark location and the expected gray-value from \eref{eq:ssam} \citep{vaananen2015generation, sarkalkan2014statistical, castro2014statistical}. We also used the anatomical shadow loss ($\mathcal{L}_{AS}$) for airways as proposed by \citet{irving20132Dxray}. Anatomical shadow compares local gray-values inside a small control area inside and outside of the projected airway surface, which can enhance airway contrast compared to dense overlapping structures such as the spinal column. Therefore, the optimal isotropic scaling, translation and shape parameters are found by minimising the equation
\begin{linenomath}
\begin{equation}
    \mathcal{L} = C_{fit} \mathcal{L}_{fit} 
    + C_{prior} \mathcal{L}_{prior} 
    + C_{g} \mathcal{L}_{g}
    + C_{AS} \mathcal{L}_{AS}
    \label{eq:ssamLoss}
\end{equation}
\end{linenomath}
where $C$ is a coefficient that controls the weight of each fitting term. 

The fit of the projected landmarks outline (`silhouette' landmarks) to the lung outline on the X-ray image is found by
\begin{linenomath}
\begin{equation}
	\mathcal{L}_{fit} = 1 - \frac{1}{\Nsl}\sum\limits_{i=1}^{\Nsl} D_i
\end{equation}
\end{linenomath}
where $\Nsl$ is the number of silhouette landmarks, $D_i$ is a function based on the distance between each silhouette landmark and the nearest X-ray image outline point. 
Silhouette landmarks were defined as landmarks where the nearest surface mesh point is shared by two faces, with one face normal pointing towards and another away from the image source. 
For example, for an anterior-posterior (frontal) chest X-ray image, the projection is normal to the $y$-direction and therefore the two faces nearest to the silhouette landmark will have positive and negative $y$ components in the normal vector.
The silhouette landmarks represent the outline of the shape normal to the projection \citep{baka20112D3D}.

\citet{baka20112D3D} normalised the outline distance to bounds of [0,1] by
\begin{linenomath}
    \begin{equation}
        D_i = \mathrm{exp} \bigg(-\frac{\mathrm{min} (||\boldsymbol{x}_{\mathrm{SSM},i}-\boldsymbol{x}_\mathrm{XR}|| )}{C}\bigg).
    \end{equation} 
\end{linenomath}
where $C=5$ was used by \citet{baka20112D3D} and also in this study. $\mathcal{L}_{fit}$ is not evaluated for the airways as the airway outline extracted from a Canny edge map is weak and unreliable due to overlapping denser regions on the X-ray image.

Improbable shapes (relative to our training set) are penalised by the Mahalanobis distance between the created shape and the mean shape \citep{baka20112D3D, ferrarini2007games} calculated by
\begin{linenomath}
\begin{equation}
\mathcal{L}_{prior} = 
\frac{1}{N_L} \sum\limits_{i=1}^{N_L}
\sqrt{
(\boldsymbol{x}_i - \overbar{\boldsymbol{x}}_i)^T \, \mathrm{cov}[\boldsymbol{x},\overbar{\boldsymbol{x}}]^{-1}\, (\boldsymbol{x}_i - \overbar{\boldsymbol{x}}_i)
},
\end{equation}
\end{linenomath}
where $N_L$ is the total number of landmarks, $\mathrm{cov}[\boldsymbol{x},\overbar{\boldsymbol{x}}]^{-1}$ is the inverse covariance matrix between $\x_i$ and $\overline{\x}_i$.

We use the appearance model to minimise the difference between the modelled gray-value ($g_{model}$) and the gray-value at the landmark location ($g_{target}$). 
This is found by
\begin{linenomath}
\begin{equation}
	\mathcal{L}_{g} = 
	\frac{1}{N_L} \sum\limits_{i=1}^{N_L}  |{g}_{model, i} - {g}_{target, i}|.
	\label{eq:grayValueFit}
\end{equation}
\end{linenomath}

Finally, the anatomical shadow loss is found for each silhouette landmark as proposed by \citet{irving20132Dxray}. 
This is done by creating two regions-of-interest (ROIs) at some distance $\pm s_{AS}$ normal to the border of the silhouette landmarks. 
Each ROI is a circle with radius $r_{AS}$, and the gray-value `inside' and `outside' of the airway projection are compared as
\begin{linenomath}
	\begin{equation}
	\mathcal{L}_{AS} = 
	\frac{1}{\Nsl} \sum\limits_{i=1}^{\Nsl} 
	\frac{ g_{inside, i} - g_{outside, i} }{ g_{outside,i} }.
	\end{equation}
\end{linenomath}
We found the optimal parameters to be $s_{AS}=20$ pixels and $r_{AS} = 14$ pixels. 

Shape parameters were optimised using the NGO (Nevergrad optimiser) algorithm \citep{meunier2021blackbox} in the Nevergrad Python library \citep{nevergrad}. This is a competence map method, which adjusts the optimisation algorithm based on the number of optimisation iterations and number of parameters optimised. We chose this method as it can therefore adjust to more modes of variation as our dataset grows in future studies. We found this gradient-free optimisation to show good robustness against local minima. The shape parameters for each mode were initialised to $\boldsymbol{b}_m=0$, and were bound to $\boldsymbol{b}_m\pm3$. 

SSAM hyperparameters in the loss function (\eref{eq:ssamLoss}) which produced the lowest maximum error in reconstructed lung space volume were found through Gaussian process regression \citep{perdikaris2016model, angelikopoulos2012bayesian}, also known as Kriging \citep{krige1951statistical}. 
We performed 100 optimisation iterations, using a subset of 11 randomly selected samples from the LUNA16 dataset. 
This gave the optimal model parameters to be $C_{fit}=0.795$, $C_{g}=0.687$ and $C_{prior} = 4.4\times10^{-4}$. 
The anatomical shadow loss coefficient was later found through a simple grid search as $C_{AS}=0.2$.

New landmarks obtained from the SSAM were converted to a surface by morphing a template mesh with template landmarks to the new set of landmarks \citep{grassi2011morphing}. 
New mesh points ($\boldsymbol{p}_{i,\mathrm{new}}$) are created from the new landmarks ($\boldsymbol{x}_{i,\mathrm{new}}$), template mesh points ($\boldsymbol{p}_{i,\mathrm{template}}$) and template landmarks ($\boldsymbol{x}_{j,\mathrm{template}}$) by
\begin{equation}
\boldsymbol{p}_{i,\mathrm{new}} = \boldsymbol{p}_{i,\mathrm{template}} 
                            + \sum_j^{N_L} k(\boldsymbol{p}_{i,\mathrm{template}}, \boldsymbol{x}_{j,\mathrm{template}}) w_j  
\label{eq:morphing}
\end{equation}
where $k$ is a kernel used for morphing the mesh and $w_j$ represents the weights controlling how much each landmark point is morphed. The weights are computed by solving \eqref{eq:morphing} for $w_j$ and setting $\boldsymbol{p}_{i,\mathrm{new}}=\boldsymbol{x}_{j,\mathrm{new}}$. We adjusted the surface morphing algorithm of \citet{grassi2011morphing} to use a Gaussian kernel, as we found it to produce a smoother deformation when morphing. The kernel is $k(a, b) = \mathrm{exp}(|| a-b||/2 \,\sigma^2)$, where $\sigma=0.3$ \citep{carr1997}.

\subsection{Segmentation from volumetric CT}
To segment lung lobes and airways from volumetric CT scans (\fref{fig:framework}d), we used convolutional neural networks \citep{lecun1995convolutional} such as the U-Net shown in \fref{fig:framework}e \citep{ronneberger2015unet, cciccek20163d}. 
We implemented the U-Net in PyTorch \citep{paszke2019pytorch}. 
The U-Net architecture was detailed by \citet{ronneberger2015unet}. 
For lung segmentation we used the pre-trained U-Net of \citet{hofmanninger2020automatic}. 
For segmenting the airways we trained a U-Net using segmentations from the LUNA16 challenge (\fref{fig:framework}e). 
Prior to training, all images and labelmaps were cropped to the extent of the airway labelmap to conserve memory and allow us to segment the entire 3D image, instead of slice-by-slice \citep{hofmanninger2020automatic}, or sliding-box approaches \citep{juarez2019joint}. 
With the aim of lowering memory consumption, we also trained an ENet (`efficient neural network') as proposed by \citet{paszke2016enet}. 
The ENet has approximately 15 times fewer parameters than the U-Net architecture \citep{comelli2021deep}, which creates lower memory consumption during training and faster processing time for 3D images \citep{paszke2016enet, lieman2017fastventricle}.

The loss function used for training all CNNs was \citep{lin2017focal, tang2019automatic}, given as
\begin{linenomath}
\begin{equation}
\mathcal{L} = \mathcal{L}_{DICE} + \mathcal{L}_{focal},
\end{equation}
\end{linenomath}
where $\mathcal{L}_{DICE}$ is the DICE loss and $\mathcal{L}_{focal}$ is the focal loss.
The DICE loss maximises the total intersection between a modelled labelmap and the ground-truth labelmap during training, and is found by
\begin{linenomath}
\begin{equation}
\mathcal{L}_{DICE} = \frac{1}{CN} \sum^C_c 
           \sum^N_i 
            \left(1 - \frac{p_{ic}g_{ic}}
                     {p_{ic}g_{ic} + (1-p_{ic})g_{ic} + p_{ic}(1-g_{ic}) }\right),
\end{equation}
\end{linenomath}
where $p_{ic}$ is the probability of a voxel $i$ belonging to class $c$ as predicted by the CNN. 
$g_{ic}$ is the ground truth value for the same voxel (which will be 1 or 0). 
This is averaged over all voxels in the image $N$ and the total number of classes $C$. 

The focal loss weights the loss more towards `harder to learn' voxels with a focusing parameter, $\gamma$. 
Focal loss was developed for images with high class imbalances, such as airway segmentations, where we found the number of voxels representing the airways to be on average $63$ times fewer than the number of background voxels. 
The focal loss was found by
\begin{linenomath}
\begin{equation}
\mathcal{L}_{focal} = -\frac{1}{CN} \sum^C_c 
           \sum^N_i 
           \alpha g_{ic} (1-p_{ic})^\gamma \log(p_{ic}),
\end{equation}
\end{linenomath}
where we set the focusing parameter $\gamma = 5$ based on a parametric study of training with $\gamma$ set to 1, 3 and 5 (Supplementary Material Figure S2). We used $\alpha=1$ as recommended by \citet{tang2019automatic}. 

To train the U-Net and ENet, we used the Adam optimiser with a batch size equal to one to minimise memory consumption. 
Data was augmented with rotations of $\pm 15 \degree$. 
We created 50 new augmented datasets per epoch. Training was stopped once the validation loss had not decreased further in the most recent 10 epochs. 
To train the U-Net and ENet models, we used a compute node with 756 GB of RAM on Heriot-Watt University high performance computing cluster ROCKS.
Training took 3 days for the U-Net (42 epochs) and 5 hours for the ENet (45 epochs). 
The ENet DICE coefficient was lower than the U-Net (0.89 compared to 0.93, Figure S3a).
The ENet had a significantly improved inference time compared to the U-Net (ENet gave 5.7 times speedup, Figure S3b).

\subsection{Distal airway generation} \label{sec:methods-distal}
To create the rest of the conducting airways not visible in CT or from our airway SSAM, we implemented a volume-filling airway generation algorithm \citep{bordas2015, tawhai2004}. 
The algorithm generates a full conducting airway tree based on information from the upper airways such as branch length and diameter.
The lung space volume was discretised into seed points with a uniform spacing found by $\Delta=(V_{lung} / N_T)^{1/3}$ where $N_T=30,000$ is the number of terminal nodes in the conducting airway tree. 

The algorithm for generating conducting airways can be found in detail elsewhere \citep{bordas2015}. 
We will briefly cover it here. 
Seed points were first clustered based on their nearest terminal branch in the airway tree. 
The centroid of each cluster was found and a splitting-plane was defined based on the centroid and the node of the parent branch. 
This splitting-plane was used to divide each seed point cluster into two more clusters. 
A new branch was generated extending towards each seed point set by 40\% of the distance to the centroid. 
When the seed point set had only one node, or the branch length was less than $2 \jwunit{\milli \metre}$, the branch is classed as a terminal bifurcation and the nearest seed point was deleted. 
This was repeated until no seed points remained in the tree. 
Once the tree was grown, a diameter was assigned to each branch segment as described in Supplementary Material Section 3.
In our CPFD simulations, the image-based airways are used until generation 3 and airway tree generated from the volume-filling algorithm \citep{bordas2015} is used to generate a surface from generations 3 to 6, similar to \citet{nousias2020avatree}.

\subsection{Deposition simulation configuration} \label{sec:deeplung-config}
To prepare the airway segmentations of the LUNA16 dataset for deposition simulations, we added a generic mouth-throat geometry obtained from a healthy adult patient \citep{ban15threedim} as done in \citet{williams2022effect} in cases where the upper airways were not imaged.
This was done by scaling the surface mesh of the healthy patient's upper airways such that the diameter at the interface between healthy patient upper airways and the patient's trachea reconstructed was the same \citep{williams2022effect}.
As jet-like flow structure produced in the upper airways has a key influence on the swirling flow observed in the trachea \citep{feng2018silico}, it is important to include at least a generic representation of the upper airways in respiratory CPFD models \citep{feng2018silico, williams2022effect}.
For the Southampton/Air Liquide dataset, segmentations of the mouth and throat were obtained by region-growing and joined to the trachea directly without scaling since they were from the same patient.

The deposition simulations in this study were performed using OpenFOAM v6 \citep{weller98tensorial} using our custom solver deepLungMPPICFoam \citep{zenodo-deepLungFoam}. 
Briefly summarised, we solved mass and momentum conservation equations for transport of an incompressible fluid (air). 
As described in \citet{williams2022effect}, our computational fluid dynamics mesh had a spacing of $\Delta=500 \micron$, which gave excellent agreement with flow through 3D printed airways \cite{ban15threedim}.
The fluid turbulence modelling is discussed in detail in \citet{williams2022pof}.
Particles were tracked in a Lagrangian approach by solving Newton's equations of motion accounting for drag and gravity, as we showed in our previous study that particle-particle interactions are not influential \citep{williams2022effect}. 
Particles were deleted when they reached an outlet, and were given a `sticking' condition when they hit the wall boundary.
The number of particles tracked was 200,000 with diameter, $d_p = 4 \jwunit{\micro m}$ as used in our previous study \citep{williams2022effect}. 
In simulations with airways from the LUNA16 dataset, particles were released from the inlet over a period of $0.1 \jwunit{s}$ with initial velocity as $10 \jwunit{m/s}$ to represent a metered-dose inhaler \citep{liu2012evaluation}. 
More information on simulation configuration for comparing to the Southampton/Air Liquide dataset is given later in this section.

To drive flow in the simulation, we model pressure at the outlet of the 3D domain using a lumped-parameter model based on a resistance-compliance circuit \citep{oakes2018airflow}. 
In this model, the pleural pressure required to drive flow during inhalation, $p_d$, is found by
\begin{equation}
p_d (t) = R_{global} Q (t) + \frac{V(t)}{C_{global}} - p_{atm},
\end{equation}
where $R_{global}$ is the global resistance of the circuit, $C_{global}$ is the global capacitance, $Q(t)$ is the flowrate as a function of time, $V(t)$ is the volume of air inhaled as a function of time, and $p_{atm}$ is the atmospheric pressure. As the flow is incompressible, there is no effect of atmospheric pressure and therefore it is dropped ($p_{atm}=0$). In simulations of inhalation in airways from the LUNA16 dataset, we model tidal inhalation where the volume of air inhaled to the lung is \citep{oakes2018airflow}
\begin{equation}
V(t) = -\frac{1}{2} \left[V_T \cos \left(2 \pi \frac{t}{T_B}\right) - V_T\right]
\end{equation}
where $V_T$ is the tidal volume and $T_B$ is the time for one breath. \citet{oakes2018airflow} estimated respiratory parameters such as $R_{global}$, $C_{global}$, $V_T$ using empirical relationships which are often based on parameters such as age, weight, height, gender. 
We used the adult parameters from \citet{oakes2018airflow}, $R_{global}=7\times 10^{-3} \jwunit{cm\, H2O\, s / mL}$ and $C_{global}=59 \jwunit{mL / cm\, H2O}$.
Pressure at the outlet is then calculated by solving the following equation at each outlet (subscript $_{(i)}$) at each time-step of our simulation:
\begin{equation}
p_{(i)} = R_{(i)} \phi_{(i)}
    + \frac{V_{(i)}}{C_{(i)}}
    + p_d.
\end{equation}
where $\phi_{(i)}$ is the flowrate (also called `flux') at the outlet computed from the fluid solver. Volume at the outlet is then found by integrating the flux in time. Local compliance and resistance at each outlet are estimated based on 
\begin{equation}
C_{(i)} = \frac{A_{(i)} \alpha_{(L)}}{A_{(L)}} C_{global}
\quad
R_{(i)} = \frac{A_{(L)}}{A_{(i)} \alpha_{(L)}} R_{global}
\label{eq:outletRandC}
\end{equation}
where the $A_{(L)}$ is the sum of areas of all outlets in each lung, $L$. $\alpha_{(L)}$ is the fraction of (static) volume of a lung to the combined volume of both lungs, which assumes that the volume change during inhalation is a factor of the lung's volume only (not accounting for localised disease). 

By defining Neumann boundary conditions for velocity at the inlets and outlets, the simulation can be unstable. Particularly, when the flow at the outlet is reversed (due to turbulence or local changes in geometry curvature) the flux at the face has a negative sign (creating positive pressure) and can cause numerical divergence \citep{esmaily2011comparison, esmaily2013modular}. To prevent backflow instabilities, we (i) set maximum pressure at each outlet to zero and minimum flux at each outlet to zero, (ii) use limiters when calculating gradients at outlet faces to prevent small negative fluxes caused by interpolation errors. We also (iii) set the maximum Courant number to be 0.1 when the flow is developing, and Courant number is set to 0.2 when the flow is developed (quantified by $0.2 < | \sin (2\pi t/T_B)|$).

Simulations of the Southampton/Air Liquide dataset modelled inhalation from a breath-actuated AKITA nebuliser \citep{conway2012controlled}.
This provided a constant inhalation flowrate of $18 \jwunit{L / min}$. 
Therefore, we assigned $V(t) = V_T \, t / T_{inhale}$, where $T_{inhale}$ is the time for one inhalation, $T_{inhale} = T_B/2$.
In the experimental study, inhalation duration was varied between $2$ and $3.3 \jwunit{s}$ (shallow and deep breathing).
However, \citet{fleming2015controlled} observed no significant difference in regional deposition when changing from shallow to deep inhalation. 
As our model cannot account for particle motion in the deep lung, we cannot model particles re-entering the central airways from the deep lung during exhalation.
Therefore, we used an inhalation duration of $2 \jwunit{s}$ for our simulations to reduce simulation clock-time, since the longer inhalation and breath-hold would only influence particle motion in the deep lung (increasing time for deposition due to diffusion or settling).
As the inhalation flowrate is steady and we could not model exhalation, we only modelled one inhalation.
This was also done by \citet{de2010validation} to compare with SPECT-CT data, although they used a steady solver which cannot capture transient flow features.

In the Southampton/Air Liquide experimental study \citep{conway2012controlled, fleming2015controlled}, the particle diameters were $d_p = 3.1$ and $6.05 \micron$.
To simplify our analysis, we used monodisperse particles as the experimental size distribution was reported as being narrow \citep{conway2012controlled}.
Simulated particles were released from a $5 \jwunit{mm}$ disk in the AKITA mouthpiece, which represented the diameter of the tubing that delivers the particles from the nebuliser.
Simulated particles were released with initial velocity of $0 \jwunit{m/s}$, since the particle timescale was small it would quickly adapt to the local fluid velocity in the mouthpiece ($\tau_p = 28 \jwunit{\micro s}$ for $d_p=3.1 \micron$ and $\tau_p = 108 \jwunit{\micro s}$ for $d_p = 6.05 \micron$). 

Simulations of flow in airways from the LUNA16 dataset were performed using ROCKS with 28 CPUs each (Intel Xeon Silver), taking between one and three days.
Simulations of flow in patient airways from the Southampton/Air Liquide dataset were performed on Oracle cloud with two bare metal compute nodes each with 36 CPUs (BM.optimised3.36, Intel Xeon Gold), taking between one and three days depending on the airway geometry.

\subsection{Model evaluation}

To evaluate our model, we compared two benchmarks.
The first aimed to verify our 2D-3D reconstruction algorithm by comparing reconstructed lung and airway morphological properties, as well as deposition results against ground truth segmentations from the LUNA16 dataset (\sref{sec:dataset}). 
The second stage aimed to validate our drug deposition modelling by comparing deposition predictions in airways segmented from CT images with our CNN to \textit{in vivo} deposition data described in \sref{sec:dataset} \citep{conway2012controlled, fleming2015controlled}. 
As our SSAM did not contain training data from the mouth-throat region, we did not test the SSAM against the Southampton/Air Liquide deposition measurements. 
The Southampton/Air Liquide experimental data contained measurements from deposition in six healthy and six asthmatic patients. 
Two experiments were performed for each patient, testing varying combinations of aerosol size, breathing pattern (shallow/deep) and inhaled gas (air compared to He/O2) giving 24 total measurements.
We omitted cases with He/O2 inhaled gas (4/24 experiments).
Images in three asthmatic patients had blockages in the upper airways due to tongue positioning, which meant we could not segment the entire upper airways and these patients were omitted (4 of remaining 20 experiments).
Additionally, as our model does not account for particle transport in the deep lung or exhalation we omitted the `shallow' inhalation in cases where shallow and deep inhalations were the independent variable compared (4 of remaining 16 cases).
Therefore, we performed 12 simulations to evaluate our model against the available experimental data \citep{conway2012controlled, fleming2015controlled}.

To validate the 2D-3D reconstruction algorithm, we reconstructed the lungs against DRRs from the LUNA16 dataset. We evaluated the reconstructed lung space volume against radiologist-verified lung segmentations from the same volumetric CT scan used to create the DRR. During the deep breathing expected when inhaling aerosol drugs, the lung space volume changes significantly. A recent modelling study \citep{koullapis2020deeplung} applied a lung expansion equal to $70\%$ relative to the initial volume to deform the deep lung's walls during inhalation. 
Bearing this in mind we aimed to have a volume error significantly below $70\%$ to produce realistic estimation of lung space volume. 
We evaluated the quality of our airway SSAM reconstruction by comparing branch diameter to the ground truth, as the diameter will influence flow characteristics such as Stokes number and Reynolds number in the central airways \citep{kleinstreuer2010review}. 
Morphological data is compared to ground truth in terms of relative error, and median, upper-quartile and 95th percentiles are assessed.

We evaluated the deposition fraction predictions in SSAM and CNN-based airways compared to simulations using the ground truth airways. 
Deposition fraction was computed as the number of particles depositing in a region (relative to the total number of inhaler particles).
Deposition predictions were assessed based on regional drug deposition fraction (right and left lung), and a local drug deposition concentration metric known as deposition enhancement factor (DEF) \citep{balashazy1999computation, longest2006childhood}. 
We calculate this using the number of particles deposited within a fixed distance ($1 \jwunit{mm}$, area $A_{conc} = \pi \, (1\jwunit{mm})^2$) of the central point of each wall face, as used by \citet{dong2019numerical}. 
This is made relative to the global deposition by
\begin{linenomath}
	\begin{equation}
	\mathrm{DEF} = \frac{ N_{p,deposit}(A_{conc})/ A_{conc}}{N_{p,deposit}(A_{tot}) / A_{tot}}, \label{eq:dosimetry}
	\end{equation}
\end{linenomath}
where $N_{p,deposit}(A)$ is the number of deposited particles in a surface area $A$.
$A_{tot}$ is the total airway surface area.
We use this instead of the sum of the deposition efficiencies as the defined areas may overlap, and this therefore prevents particles being counted multiple times towards the global average. 
To minimise differences in total number of deposited particles or total surface area interfering with quantitative comparison between DEF in the ground truth and SSAM or CNN, we used the ground truth denominator from \eref{eq:dosimetry} in DEF calculations in the SSAM and CNN \citep{dong2019ultrafine}.  
Finally, we validated our framework by quantifying simulated regional deposition (left lung, right lung and mouth-throat airways) against experimental values reported by \citet{conway2012controlled} and \citet{fleming2015controlled}.
Relative errors are expressed in terms of the median, upper quartile and maximum. 
Correlation between experimental and simulated deposition data is also assessed using the concordance correlation coefficient.
All statistical analyses were performed in Python 3.8.

\section{Results}
\subsection{Morphological assessment on LUNA16 dataset}
\begin{figure}
\centering
    \includegraphics{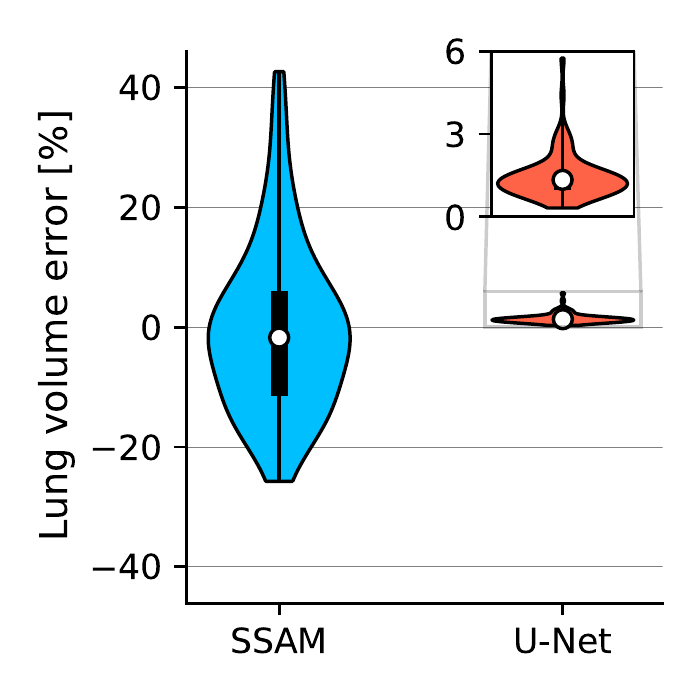}
\caption[Lung space volume error on LUNA16 dataset]{Lung space volume error from SSAM and U-Net reconstructions of patients from LUNA16 dataset. Inset shows U-Net lung space volume error, where the U-Net results were produced using the pretrained U-Net from \citet{hofmanninger2020automatic}.\label{fig:luna16VolErr}}
\end{figure}

We found our SSAM to yield a relative lung space volume error of 9.9\% median and a 95th percentile of 34.3\% on the LUNA16 dataset (\fref{fig:luna16VolErr}a). 
The maximum error for the right lung was 33.64\% and 41.66\% for the left lung. 
The concordance correlation coefficient between SSAM lung volume and ground truth lung volume was 0.884 and 0.893, respectively for one and two DRRs used for SSAM fitting (Figure S4). 
The total volume error (averaged over left and right lungs) had a median value of 9.8\% and a 95th percentile of 26.1\%. 
Larger error in the left lung may be due to the presence of the heart interfering with the edge-map, which would cause a poor fit at the inner side of the lung.
In contrast, the absolute relative error for the U-Net lung segmentation gave a median value of 1.35\% and a 95th percentile of 3\% (\fref{fig:luna16VolErr}).
The U-Net is well-suited to segmenting lungs from CT, due to the high-contrast and well-resolved boundary between dark (air-filled) lung parenchyma and surrounding soft and hard-tissue.
The SSAM lung reconstruction yielded a larger maximum error due to the poor contrast between lung and surrounding tissues on some radiographs.

\begin{figure}
\centering
\begin{tabular}{ccc}
    (a) largest error & (b) second largest error & (c) lowest error \\
    \includegraphics[width=0.3\linewidth]{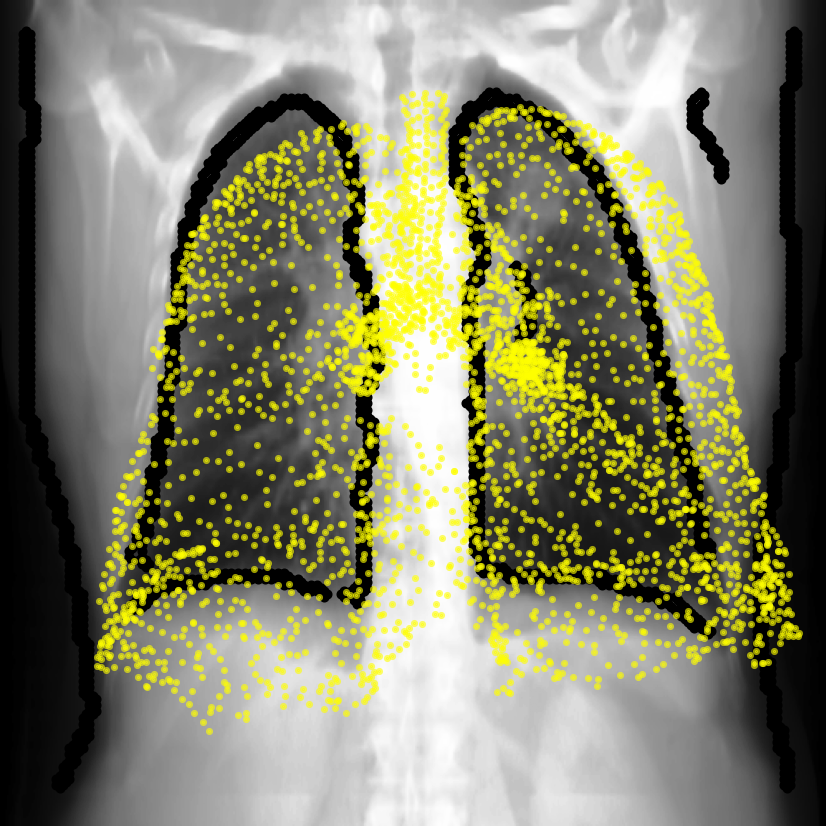} &
    \includegraphics[width=0.3\linewidth]{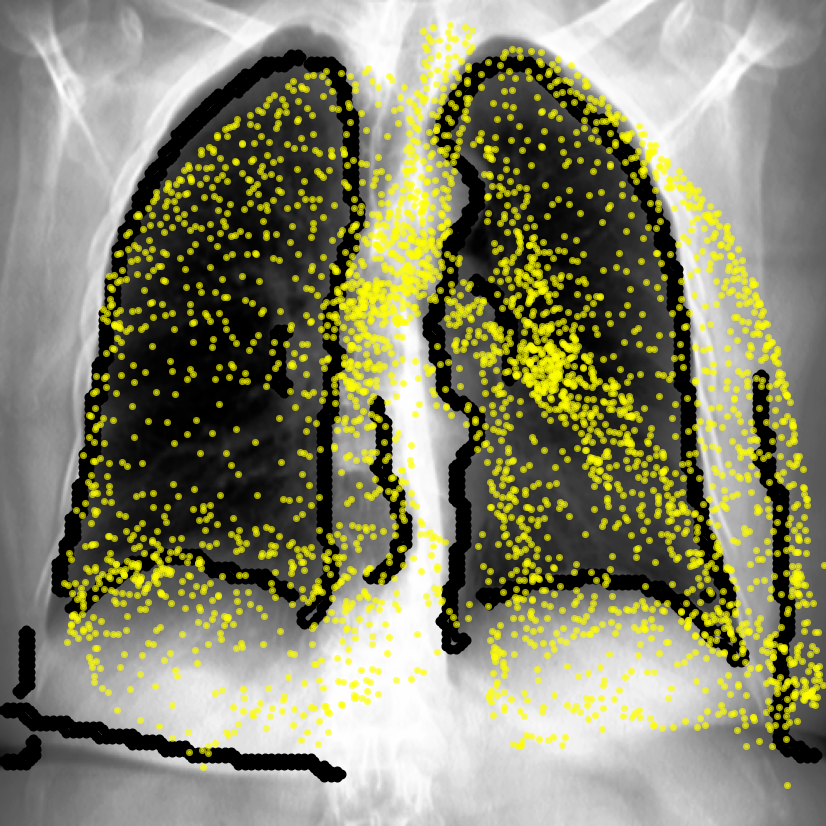} &
    \includegraphics[width=0.3\linewidth]{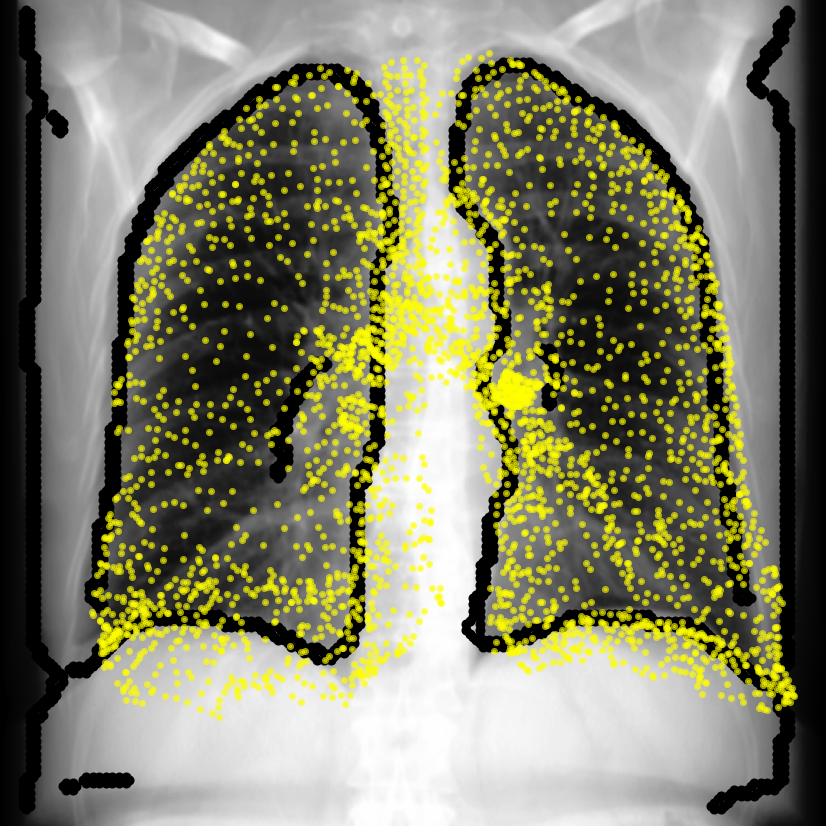}
\end{tabular}
\caption[Comparison of SSAM outline fit in outlier cases]{Example of fitted SSAM compared to projection and lung outline for two largest lung space volume errors. 
Yellow markers are output landmarks from the SSAM, and black markers represent the X-ray edges detected by the Canny edge map described in \sref{sec:preproc}.
Panels show (a) the largest error, (b) second largest error and (c) the lowest error. 
In (a) the right lung error was 31\% and left lung error was 42.6\%. 
In (b) the right lung error was 15.5\% and left lung error was 38.3\%.
In (c) the right lung error was 1.8\% and left lung error was 0.06\%.}
\label{fig:lungVolOutliers}
\end{figure}

In \fref{fig:lungVolOutliers} we compared the output SSAM landmarks against the input X-ray image and lung outline map in two outlier cases and the best case (in terms of lung space volume error, \fref{fig:luna16VolErr}).
In both cases, we observed that the outer edge of the left lung has incorrectly fitted to a spurious edge introduced to the lung edge map. 
In the largest error case (\fref{fig:lungVolOutliers}a), the lower corner of the SSAM landmark point cloud has fitted to the outline of the patient's torso.
In the case with the second largest error (\fref{fig:lungVolOutliers}b), the left lung outer edge and right lung lower edge have mistakenly fitted to an edge in the soft tissue.
In the lowest error case (\fref{fig:lungVolOutliers}c), there is a spurious edge for the outline of the patient's torso that slightly interferes with the outer edge of the patient's right lung.
However, this is located very near the lung itself as the patient has a narrow torso, which therefore appears to cause a minor influence on the SSAM fitting.

\begin{figure}
\centering
\includegraphics{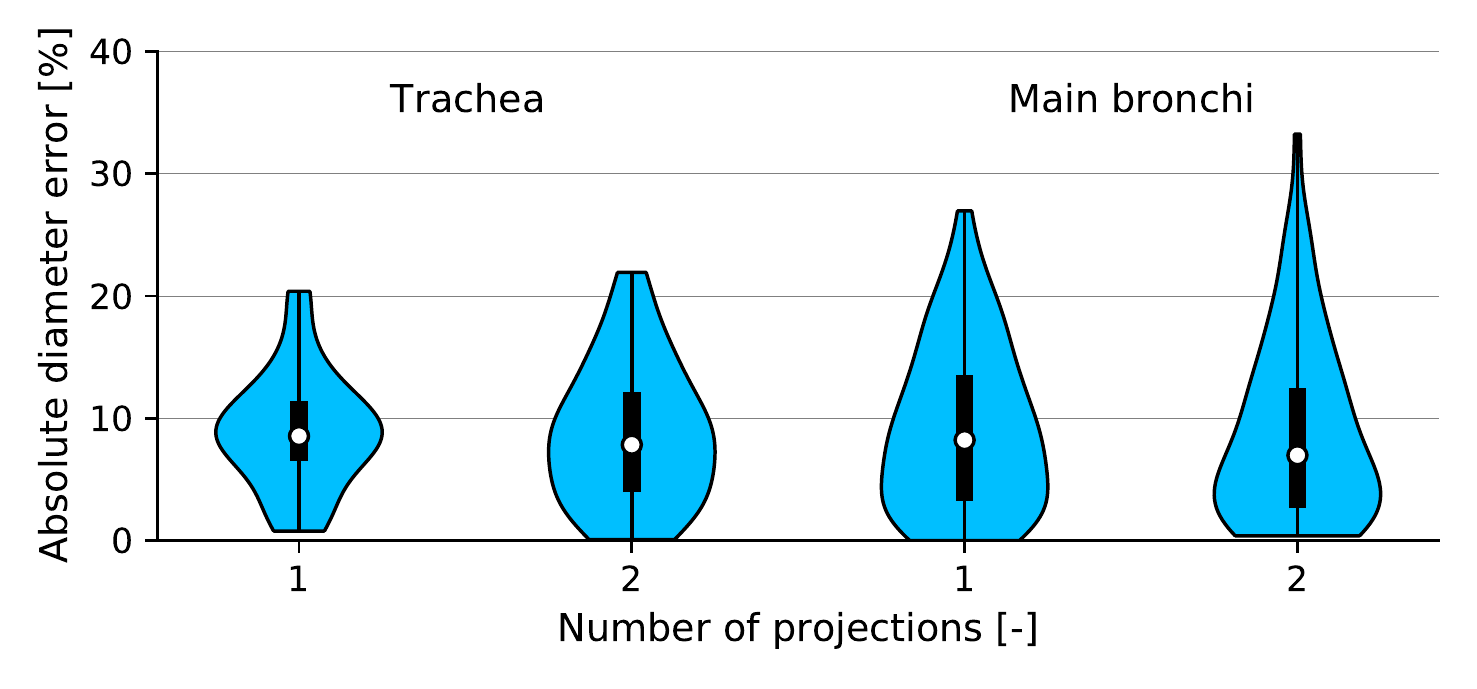}
\caption[Airway diameter error on LUNA16 dataset]{Comparison of airway diameter errors predicted by our SSAM compared to the ground truth segmentations. 
We show the influence of including an additional projection on diameter error in the trachea and main bronchi. }
\label{fig:luna16Airway}
\end{figure}

Diameter predictions did appear to be sensitive to outliers, as the maximum diameter error was above 30\% in the main bronchi (\fref{fig:luna16Airway}).
We compare airway diameter error with a single DRR (anterior-posterior plane projection) and two DRRs (anterior-posterior and sagittal plane projections).
The maximum trachea diameter error was 20.4\% for one projection and 21.9\% for two projections, which shows the trachea is less sensitive to outliers than the main bronchi due to higher visibility on an X-ray image.
The absolute trachea diameter error had a median value of 8.5\% and 7.8\% for one and two projections, respectively.
The main bronchi maximum error was 26.9\% and 33.2\% for one and two projections, and the upper quartile error was 13.6\% and 12.4\%.
The concordance correlation coefficient was slightly higher for two projections than one (0.71 compared to 0.657, Figure S5).
There was no statistically significant difference between the absolute error when comparing SSAM results with one and two projections (Wilcoxon signed-rank test $p=0.6$ and $p=0.69$ for the trachea and main bronchi, respectively).

\begin{figure}
\centering
\begin{tabular}{c c}
    \includegraphics[trim={0 0 0 7mm},clip]{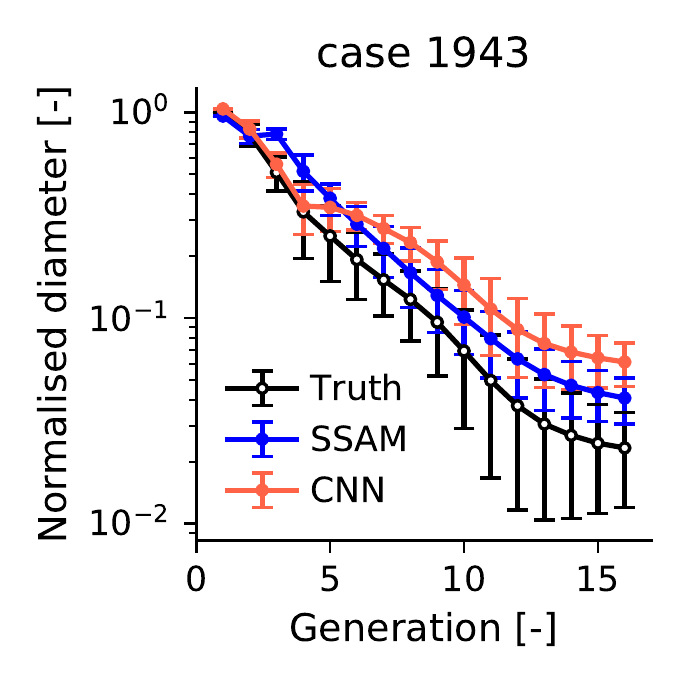} & 
    \includegraphics[trim={0 0 0 7mm},clip]{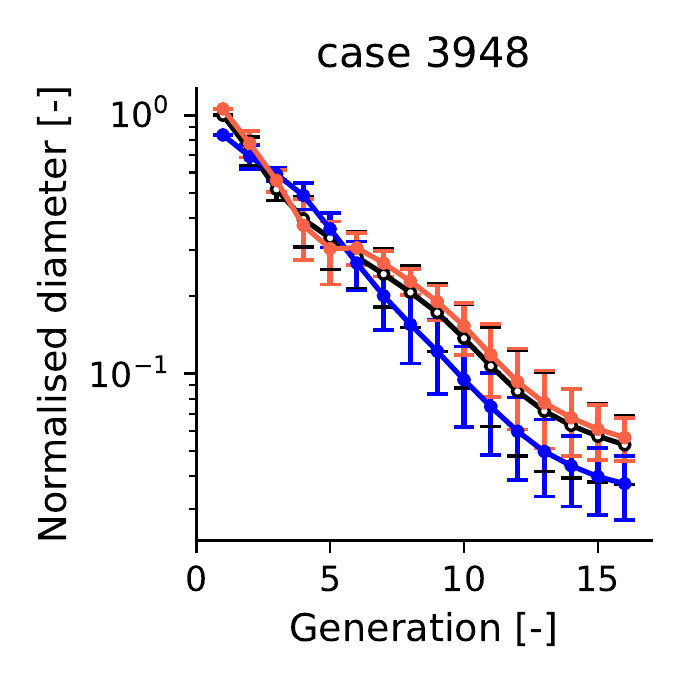} \\ 
\end{tabular}
\caption{Generated airway diameter for our image-processing modalities on two cases from the LUNA16 dataset. Markers show the normalised mean diameter per generation. Error bars show $\pm 1 S.D.$ where $S.D.$ is the standard deviation.}
\label{fig:genAirwayDiaCNNvsSSAM}
\end{figure}

To evaluate the effect of imaging error propagation into distal airway diameters, we evaluated the diameter generated from the distal airway generation algorithm from the ground truth compared to the SSAM and CNN (\fref{fig:genAirwayDiaCNNvsSSAM}).
Here, the ground truth refers to results from the airway generation algorithm using the central airway tree from the ground truth segmentations.
Due to the dependence on the diameter in the central airways, the CNN also shows mixed performance.
The median absolute error in mean airway diameter per generation was 24.2\% (similar to the SSAM with parent model, Figure S6).
Additionally, the maximum absolute error was 77.6\%, compared to 76\% for the SSAM.
The CNN airways show improved predictions of the diameter standard deviation (\fref{fig:genAirwayDiaCNNvsSSAM}), as the maximum error in standard deviation was 125.5\% for the SSAM and 69.6\% for the CNN.

\subsection{Deposition analysis}
In this section, we first compare deposition predictions from CPFD simulations in the LUNA16 dataset. We compare predicted deposition in ground truth airways with deposition predictions in airways obtained from SSAM and CNN segmentations. Following this, we then compare our simulated deposition from airways segmented by the CNN with experimental results from the Southampton/Air Liquide dataset.

\begin{figure}
\centering
\begin{tabular}{cc}
    (a) & (b)  \\
    \includegraphics{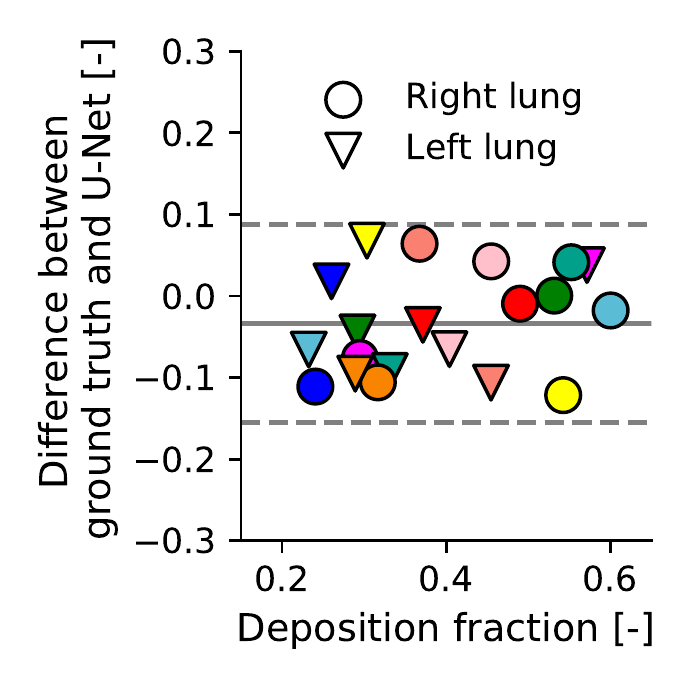} & \includegraphics{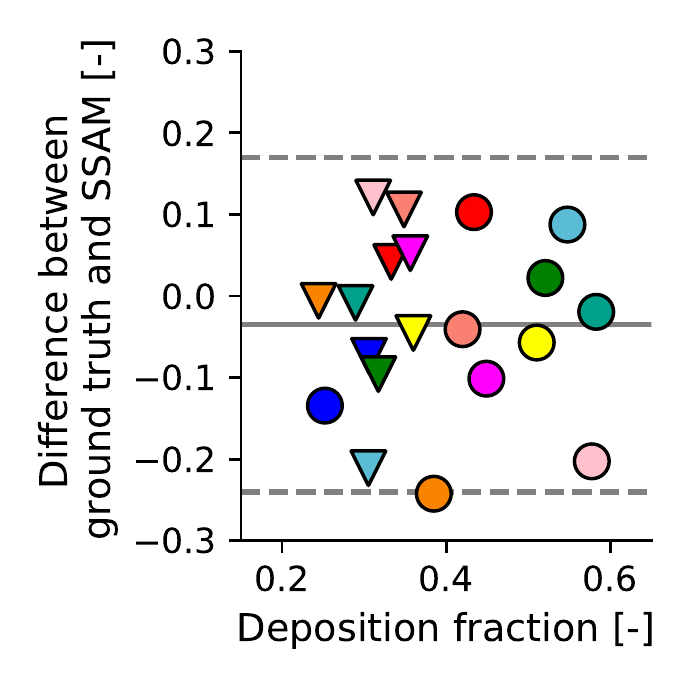}
\end{tabular}
\caption{Bland-Altman plot comparing drug deposition in each lung predicted in ground truth airways, compared to airways reconstructed by our SSAM. Number of patients in sample was 9. Each marker represents one lung, and each colour represents a different patient.}
\label{fig:luna16DepositionAll}
\end{figure}

Errors in deposition fraction from simulations of flow in airways from the LUNA16 dataset are shown in \fref{fig:luna16DepositionAll}.
Comparing simulations in the CNN with the ground truth, the median absolute error was 6.5\%, with a 95\% confidence interval of $-3.3 \pm 12.1\%$ (\fref{fig:luna16DepositionAll}a). 
The Pearson correlation coefficient between ground truth and CNN deposition results was 0.88.
Deposition in the SSAM airway had a similar median absolute error as the CNN deposition (8\%, \fref{fig:luna16DepositionAll}b).
SSAM deposition error was more widely distributed, as the 95\% confidence interval was $-3.5 \pm 20.5\%$.
The Pearson correlation coefficient between deposition in the SSAM with the ground truth was 0.6.
The error range for deposition in the SSAM was similar to diameter error (\fref{fig:luna16Airway}) which had a confidence interval of $-2.1 \pm 21.7\%$.

\begin{figure}
\centering
\includegraphics{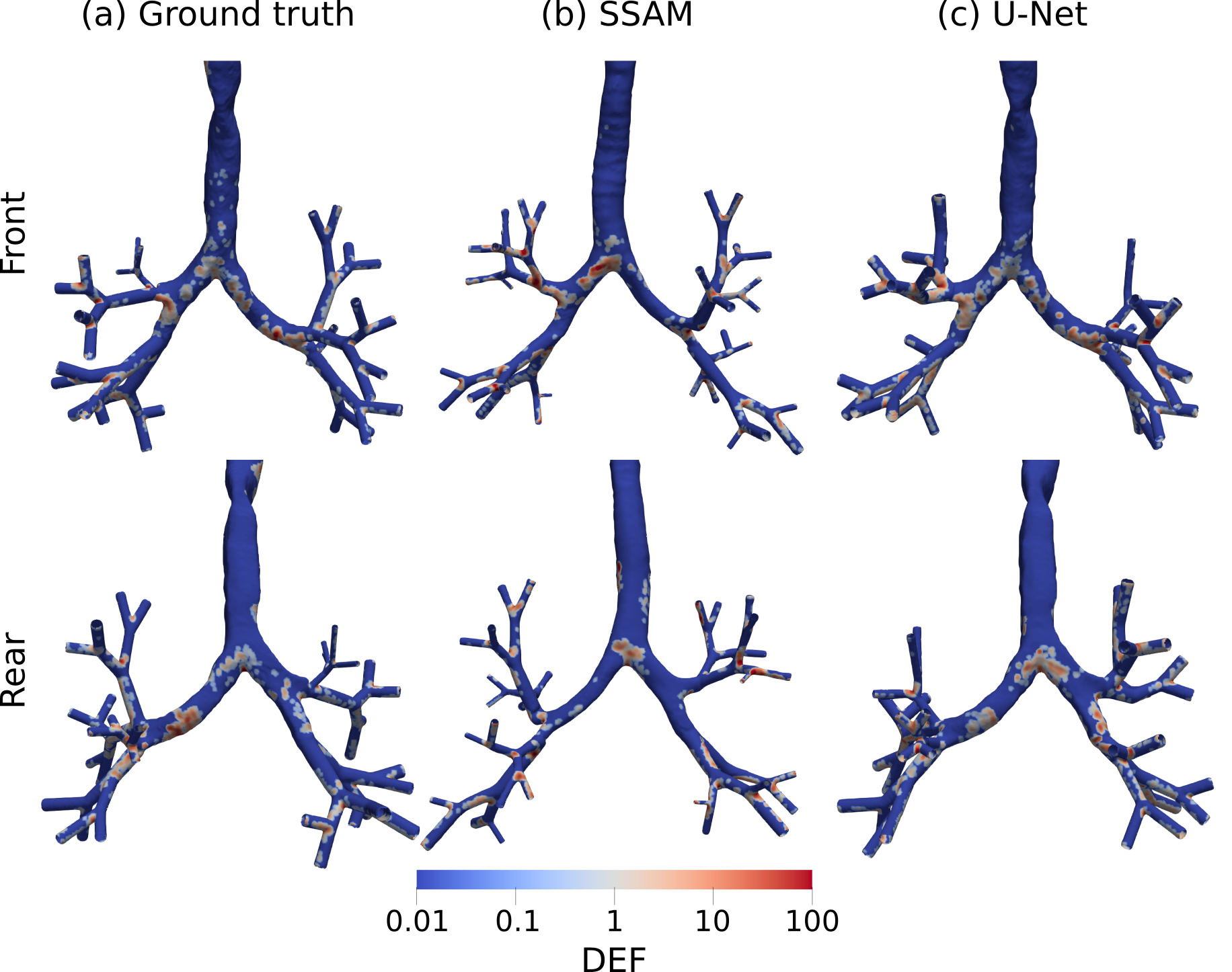}
\caption[Deposition concentration for different image-processing methods compared to ground truth airways]{Deposition enhancement factor for simulations in airways constructed using the developed SSAM and U-Net, compared to simulations in a ground truth segmentation. 
Columns show (left) ground truth, (middle) SSAM, and (right) U-Net segmented airways. 
Rows show (top) front and (bottom) rear view of the airways.}
\label{fig:luna16DepositionDEF}
\end{figure}

To evaluate the effect of SSAM reconstruction on local deposition hotspots, we computed the deposition enhancement factor (DEF).
The deposition hotspots are highly sensitive to local flow and geometry \citep{farghadan2020particle, williams2022effect}, which itself is susceptible to geometric differences by image acquisition and image processing \citep{macdonald2020experimental}.
We can see that even for the case with the lowest deposition error (\fref{fig:luna16DepositionDEF}), the deposition hotspots in the SSAM do not match the ground truth. 
Particularly, in the ground truth and CNN segmented airways (\fref{fig:luna16DepositionDEF}a,c), the drug is noticably more dispersed than in the SSAM (\fref{fig:luna16DepositionDEF}b).
Less dispersed hotspots in the SSAM is likely due to additional turbulence generated in the CT-based airways which is not present in the SSAM, since it contains a reduced number of principal components.
A lower number of principal components essentially simplifies the geometry and removes intricate details that may onset turbulence such as cartiligenous rings in the trachea \citep{zhang2005measurement, russo2008effects}.

Deposition hotspots in the CNN-based airways appear qualitatively similar to the ground truth (\fref{fig:luna16DepositionDEF}a,c). 
The surface area with DEF $>1$ at the first bifurcation in the ground truth is 17\% larger than the CNN.
However, the intensity of the hotspots is significantly different since the average DEF at the first bifurcation was 2.49 for the ground truth (maximum value 400), compared to 0.93 for the CNN (maximum value 225).
At the first bifurcation, the mean DEF in our SSAM was 1.06 (maximum value 225). 
The surface area with DEF $> 1$ at the first bifurcation was 2.16x larger in the ground truth than the SSAM.
Interestingly, the mean DEF at the first bifurcation in the SSAM and U-Net are similar (1.06 compared to 0.93 for SSAM and U-Net, respectively). 
However, there is a significant difference in the surface area covered by the deposition hotspots here  (\fref{fig:luna16DepositionDEF}).

\begin{figure}
\centering
\begin{tabular}{cc}
    (a) & (b) \\
    \includegraphics{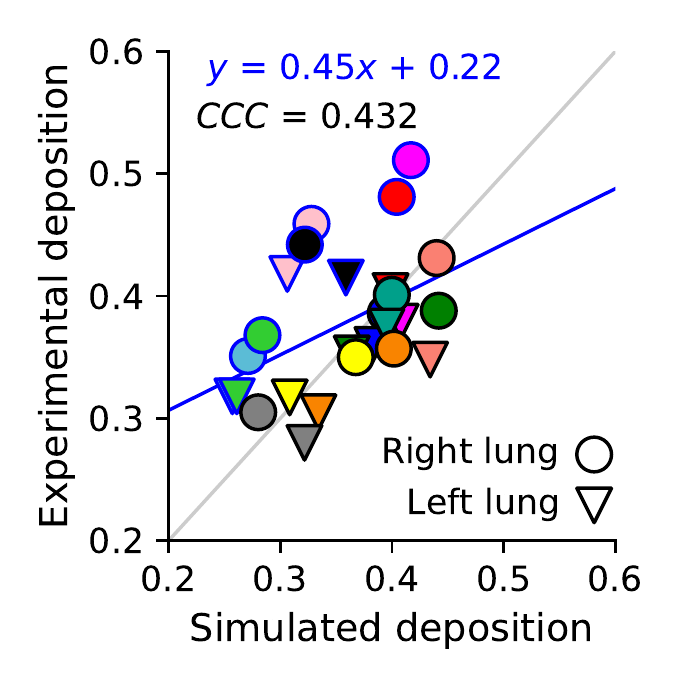} &
    \includegraphics{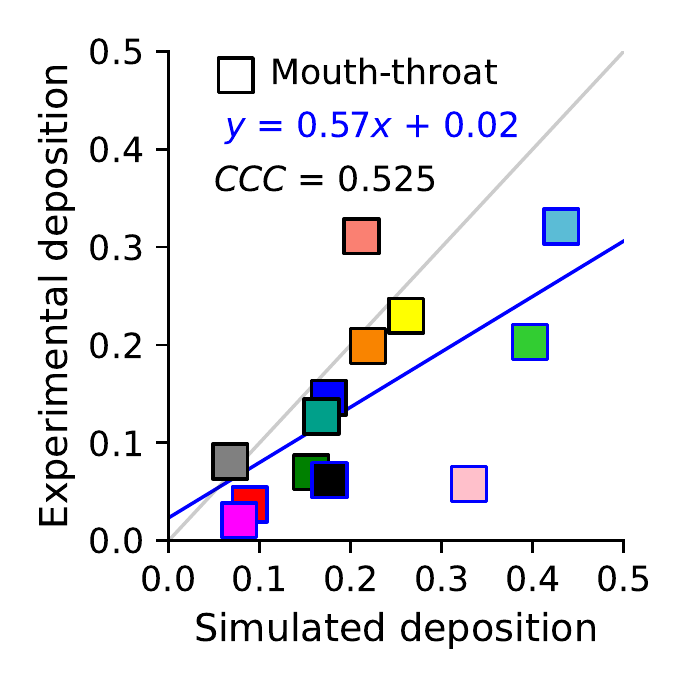}
\end{tabular}
\\(c)\\
\includegraphics{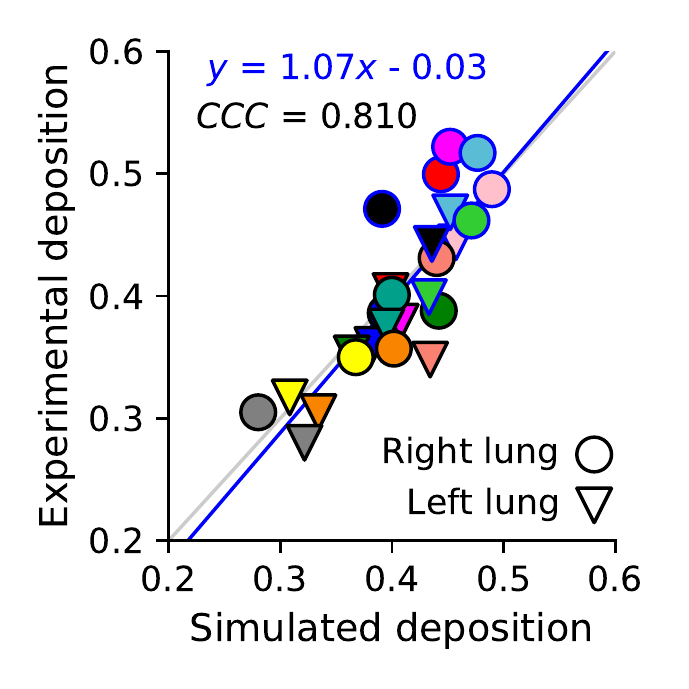}
\caption[Comparison of simulated deposition against experimental data]{Comparison of simulated deposition against experimental measurements from \citet{conway2012controlled}.
Panels show (a) right and left lung deposition, (b) mouth-throat deposition, and (c) the partially corrected right and left lung deposition.
Cases which are adjusted by correcting over-predicted mouth-throat deposition are shown with a blue border on the marker.
$CCC$ is the concordance correlation coefficient.
The linear regression lines shown have (a) $r^2=0.213$, (b) $r^2=0.36$, (c) $r^2=0.699$.
The cases requiring no manual cleanup of the glottis have grey and yellow markers.
}
\label{fig:sotonDepo}
\end{figure}

Finally, we performed simulations of nebulised aerosol inhalation in healthy and asthmatic patient cases of the Southampton/Air Liquide dataset \citep{conway2012controlled, fleming2015controlled}. 
The maximum deposition absolute error averaged over both lungs was 12.1\%, which belonged to the case with most obstructed upper airway images shown in Figure S1a (case H02 with small particles). 
The median, 75th percentile and maximum absolute error was 4.9\%, 7.9\% 13.1\%, respectively.
The concordance correlation coefficient between simulated and experimental lung deposition measurements was 0.43. 
The mouth-throat deposition had a concordance correlation coefficient of 0.525 (\fref{fig:sotonDepo}b).
The maximum mouth-throat deposition error was also in case H02 with small particles (27.2\%).
The linear best-fit line shows that the cases with high error in the mouth-throat region causes the distribution to depart from the 45 {\degree} line (\fref{fig:sotonDepo}a,b).
We observed that mouth-throat deposition was best in the two cases where the glottis did not require manual cleanup (grey and yellow markers in \fref{fig:sotonDepo}b).
The error in these cases was 1.3\% and 3.1\%.
The median and 75th percentile mouth-throat deposition errors were 5.7\% and 11\%, respectively.
These results show that our models agree well with \textit{in vivo} deposition data, particularly in the absence of imaging artefacts (\fref{fig:sotonDepo}).

In an attempt to understand how our simulations would perform in a perfect case with no imaging difficulties in the upper airways, we partially corrected the simulation and experimental data by normalising the deposition fraction by $1-DF_{mouth}$, where $DF_{mouth}$ is the deposition fraction in the mouth (\fref{fig:sotonDepo}c).
This makes deposition relative to number of particles entering the trachea, rather than the total number of particles entering the mouth.
We apply this only to cases where the experimental lung deposition fraction was 5\% larger than the simulated lung deposition (only outlier cases above the 45{\degree} line), as these were cases where the mouth-throat deposition was severly over-estimated (for example,  green markers in \fref{fig:sotonDepo}).
Once these problematic cases were corrected, we see an improvement in the correlation between experimental and simulated data, as the concordance correlation coefficient increases from 0.432 to 0.810 (\fref{fig:sotonDepo}a,c).
This can be seen quantitatively by the fitted linear regression line nearly matching the 45{\degree} line.
The resultant $r^2$ value is 3.3 times larger, at a value of $r^2=0.699$.
The median, 75th percentile and maximum errors all decreased to 2.12\%, 4.14\% and 8.6\%, respectively.
This idealised test shows that the physical model is working well, but manually cleaning the segmentations around the glottis can introduce uncertainty that lowers correlation with experimental data (\fref{fig:sotonDepo}a,c).
Additionally, this idealised test points to an avenue for future development of image-processing tools for this region.

\section{Discussion} 
In this study, we aimed to develop an image-processing framework to automatically generate personalised computational respiratory systems from volumetric CT and planar X-ray image(s).
To that end, we developed a statistical shape and appearance model to parameterise the variation in 3D lung and airway shapes across a population.
An algorithm was proposed to reconstruct 3D lung and airway geometries, by iteratively adjusting shape parameters based on a given X-ray image.
We also trained a convolutional neural network (CNN) based on the U-Net and ENet architectures \citep{ronneberger2015unet, paszke2016enet}, which allow us to automatically segment airways from volumetric CT scans.
We evaluated the ability of our proposed framework to reproduce morphological properties and deposition results that were obtained from the ground truth airway/lung geometries.
The CNN-based airways showed good agreement with the ground truth on all metrics.
Our SSAM showed good agreement with the ground truth, although it was sensitive to outliers in predictions of morphological properties due to the low amount of information available in a planar X-ray image.
Finally, we aimed to validate our proposed framework against experimental \textit{in vivo} deposition measurements \citep{conway2012controlled, fleming2015controlled}.
After cleaning the upper airway segmentations to account for glottis motion artefacts, we obtained good agreement with the experimental data.

Despite imaging uncertainties in the upper airway geometry, we obtained good agreement with experimental deposition measurements (median deposition error 4.9\%, \fref{fig:sotonDepo}).
Available \textit{in vivo} experimental studies of drug deposition are limited due to the difficult and expensive nature of the experiments.
Therefore, most modelling studies with validation compare to \textit{in vitro} experiments, such as flow or deposition measurements in airway casts \citep{williams2022effect, koullapis18regional, holbrook13validating, longest2012comparing}.
Alternatively, some CPFD validation studies have used varying forms of experimental \textit{in vivo} data \citep{tian2015validating, de2010validation, oakes2013regional, oakes2015distribution}.
\citet{tian2015validating} compared their geometrically-simplified model to 2D scintigraphy data at varying particle sizes.
\citet{de2010validation} compared patient-specific models with SPECT-CT data, but only for tracer particles ($d_p \approx 1 \micron$).
Additionally, their outlet boundary conditions uses a fixed flowrate based on the regional lung space volume change from CT images at full inhalation and exhalation.
In contrast, the outlet pressure in our study is approximated with a model based on lung space volume fraction, resistance and compliance (\sref{sec:deeplung-config}). 
Our approach has the benefit that it requires less radiation (only one CT or X-ray image required) to model regional ventilation to each lung.
Other \textit{in vivo} validation studies have compared to 3D deposition measurements in rats \citep{oakes2013regional, oakes2015distribution}, although the Reynolds number and Stokes number will change in these cases since the airway diameter is smaller than humans.
Therefore, by validating our full patient-specific modelling and imaging approach against 3D \textit{in vivo} data at varying particle sizes, our work represents a key development in advancing towards incorporating deposition models into clinical care.

Morphological and deposition results produced by the developed SSAM had a wide distribution, as the SSAM was sensitive to outliers. 
We observed that outliers occurred in cases where the outline of the lung on the X-ray image was noisy or low quality (\fref{fig:lungVolOutliers}).
However, the SSAM largely provided good agreement with the ground truth, as median absolute error was under 10\% for the lung space volume (\fref{fig:luna16VolErr}), airway diameter (\fref{fig:luna16Airway}), and lung deposition (\fref{fig:luna16DepositionAll}b).
We also observed that the maximum error for airway diameter (32\%) was less than the maximum error for regional deposition (24\%), which shows that morphological errors in the SSAM do not propagate into significantly worse regional deposition results despite the nonlinear influence of diameter error on local velocity, Reynolds number and Stokes number on deposition.

When assessing local deposition hotspots, we observed significantly different deposition patterns in the SSAM compared to the ground truth (\fref{fig:luna16DepositionDEF}a,b).
Therefore, for exact predictions of deposition hotspots, the SSAM is not a suitable approach and a CNN segmentation from CT should be used.
This occurs since hotspots are highly influenced by near-wall flow structures. 
This shows the complex geometrical features of the airways that create secondary flows responsible for deposition cannot be reconstructed from the limited information available in a planar X-ray image.
Our SSAM shows similar performance to the SSAM developed by \citet{vaananen2015generation} to reconstruct femora from 2D radiographs, as the average errors were below 10\%, but the SSAM had difficulty predicting local morphological properties.
The quality of our reconstruction from 2D images may be improved by using a neural network (transformer) architecture to find the optimal shape parameters ($\boldsymbol{b}$) from the X-ray image, as this approach was shown by \citet{shen2019patient} to reconstruct a full volumetric CT from a series of X-ray images.

By analysing the lung space volume error outliers, we observed that the input lung edge map appeared to create some difficulties due to spuriously introduced edges (\fref{fig:lungVolOutliers}).
\citet{li2016automatic} proposed an approach to extract lung-field outlines from chest X-rays using SSAMs.
Using a SSAM instead of a general edge detection algorithm constrains the lung outline with the SSAM modes, and ensures no non-lung edges are highlighted to potentially interfere with the fitting.
Alternatively, fitting of the SSAM could be improved by replacing the global loss function parameters (model coefficients, $C$ in \eref{eq:ssamLoss}) with patient-specific values. 
This could be obtained by determining the optimal parameters for each patient in the training database, through the same Gaussian process regression described in \sref{sec:methods-ssam}.
The input chest X-ray may then be passed through a convolutional neural network with four output neurons (one for each $C$ in \eref{eq:ssamLoss}).
These minor alterations may improve fitting the loss function given in \eref{eq:ssamLoss} and yield a more robust SSAM algorithm.

Error between experimental and simulated deposition in CNN-segmented airways in the LUNA16 dataset gave a 95\% confidence interval of $-3.3\pm 12.1\%$ (\fref{fig:luna16DepositionAll}a).
This is likely due to errors in distal airway diameter (\fref{fig:genAirwayDiaCNNvsSSAM}) that influenced deposition in the distal airways and particles reaching the deep lung. 
The distal airway diameter may also have influenced CNN-segmented airway deposition through changing the outlet diameter and the computed outlet boundary condition through \eref{eq:outletRandC}.
The ability of the CNN to segment distal airways was limited by the low resolution of LUNA16 datasets (mean slice thickness $1.6 \jwunit{mm}$), which is much lower than the high resolution CT used in the Southampton/Air Liquide dataset (mean slice thickness $0.5 \jwunit{mm}$) \citep{conway2012controlled, fleming2015controlled}.
We are also limited by the quality of ground truth obtained by our semi-automatic algorithm which was verified by a radiologist to have no leakage, but suffered from a limited number of branches segmented.
This was due to the low image resolution, which can create significant difficulty in obtaining a full airway tree without leakage \citep{tschirren2005intrathoracic, lo2012extraction}.
The CNN training could be improved to mitigate this by masking voxels belonging to trachea and main bronchi, which forces the model to focus on learning to segment small airways \citep{garcia2021automatic}.

To move towards incorporating this framework in a clinical decision-making workflow, the bottleneck of simulation time is the primary obstacle to overcome.
Clock-time for simulations in this study ranged between 1-7 days, which is not suitable for clinical use.
To deliver real-time prediction of clinical quantities of interest, geometric parameters (such as SSM weights) and flow variables can be used to fit machine learning models to simulation data to provide predictions of deposition within minutes \citep{hoeijmakers2020combining, hoeijmakers2021uncertainty, su2020generatingWSS}.
\citet{morales2021deep} showed that wall-shear stress could be predicted from patient-specific vascular meshes with minimal pre-processing using geometric deep learning, which has been developed for analysing non-Euclidean geometries such as graphs or meshes \citep{bronstein2017geometric}.
This approach could be applied to the respiratory system to evaluate properties such as deposition hotspots (\fref{fig:luna16DepositionDEF}) based on a mesh obtained with our segmentation algorithm, without the additional step of fitting to a SSM as done by \citet{hoeijmakers2020combining} for the aortic valve.
An alternative to using classic purely data-driven machine learning models to predict deposition is to constrain the neural network fitting with some physical constraints such as conservation of mass and momentum \citep{raissi2018hidden, raissi2019physics, alber2019integrating, karniadakis2021physics}.
This approach generates full velocity and pressure fields based on the physical constraints and training data from similar flow configurations.
However, it is unclear if the physics-informed neural network approach would work in the geometrically complex airway tree, since its application has been limited to simple flow configurations such as bluff body flow, lid-driven cavity or pipe flow \citep{raissi2018hidden, yin2021noninvasive, jagtap2020conservative}.
Our developed imaging and simulation framework could be used to generate training data for machine learning approaches, as well as to benefit from the speed improvements created by machine learning for CFD.
Combining these modelling approaches is crucial to feasibly incorporate patient-specific models in clinical practice.

\subsection{Limitations}
A limitation of this study was the difficulty in capturing the upper airway anatomy (Supplementary material Section 1). 
The glottis can dilate and narrow during inhalation and exhalation, respectively \citep{brancatisano1983vocalcords, scheinherr2015realistic}, which created difficulties in capturing CT images of the glottis \citep{conway2012controlled}. 
Changes in the shape and area of the airways here can cause increased deposition in the glottis \citep{zhao2020glottis}, but also downstream due to the flow separation at the expansion which creates an unsteady wake whose structure will likely vary based on patient anatomy \citep{luo1996numerical, cisonni2010experimental, bhardwaj2022glottal}.
Therefore, our attempt at cleaning up the glottis cannot perfectly represent instantaneous patient-specific flow patterns observed \textit{in vivo}.
Despite this, we observed good agreement with experimental lung and mouth-throat deposition measurements (\fref{fig:sotonDepo}).
Future efforts should be focussed towards combining our developments with dynamic upper airway models \citep{zhao2020glottis, bhardwaj2022glottal}, to enable modelling the complex biomechanical fluid-structure interaction of respiration.
However, to perform fluid-structure interaction simulations of these cases still requires a reliable baseline airway model, which requires manual cleanup of outlier cases shown in \fref{fig:sotonDepo}.
An improved approach to manually cleaning the glottis segmentations may be to determine the missing airway structure based on statistical shape models, as done for segmenting obstructed bronchi by \citet{irving2011segmentation}.
Therefore, a template mesh could be morphed to approximate the glottis shape based on the shape of the nearby upstream and downstream airways (the throat and trachea).
Combining this improved segmentation cleanup approach with dynamic mesh simulations in the upper airways is likely to produce a robust approach for modelling upper airway dynamics.

An additional limitation is the use of a healthy patient mouth-throat geometry \citep{ban15threedim} as a representative generic geometry for the upper airways in the LUNA16 dataset.
Due to unavailable imaging data above the trachea in most clinical CT datasets, it was not possible to segment this from CT with our CNN, or include it as part of our SSAM to reconstruct from X-ray images.
This of course limits the understanding of how well our imaging approach can capture the upper airways and flow patterns specific to a patient.
The effect of this would be negligible on the CNN segmentation, as the upper airways are easily segmented with a region-growing approach \citep{garcia2021automatic}.
It is unclear how well the SSAM could reconstruct the mouth-throat from a patient X-ray image, as there is no existing SSM that shows the main modes of variance in this part of the airways. 
However, literature agrees that the glottis cross-sectional area is the most significant upper airway morphological parameter influencing deposition \citep{xi2016parametric, feng2018silico}.
Moreover, realistic mouth-throat geometries were shown to give similar regional deposition predictions as a simplified mouth-throat geometry with an elliptical cross-section when $d_p \leq 10 \micron$ \citep{xi2016parametric}.
This suggests that it would not be essential for a SSAM to capture the exact shape of the mouth-throat, but an accurate prediction of the glottal cross sectional area is required. 
With a good estimation of glottal area, the SSAM could produce deposition that agrees with the deposition as shown for our generic healthy patient mouth-throat (\fref{fig:luna16DepositionAll}).

Our deep lung model was a 0D approximation of deep lung mechanics \citep{oakes2015distribution, oakes2018airflow}.
Using this 0D model as an outlet boundary condition for the 3D domain (3D-0D model) has been shown to accurately predict lobar deposition in healthy and emphysematous rats \citep{oakes2015distribution}.
In this 0D model, particles reaching the outlets are deleted and classified as depositing in the deep lung.
\citet{oakes2017aerosol} developed a 1D model for fluid and particle transport in distal airways (3D-1D model), which tracks particle concentration in the distal airways. 
This 1D formulation showed minor improvements in agreement with experimental lobar deposition measurements over the 0D formulation \citep{oakes2017aerosol}.
A key benefit of the 1D formulation is that particles can re-enter the 3D domain during exhalation as it is known how many particles are still floating in the distal lung \citep{oakes2017aerosol}.
An inability to model particles re-entering the domain during exhalation is a limitation of our model, as we cannot compare the fraction of drug exhaled with experimental measurements \citep{conway2012controlled, fleming2015controlled}.
In the majority of experiments, the exhaled fraction was below 10\%, meaning this has only a minor influence on the validity of our model for inhalation studies.
Incorporating a 1D model for particle and flow transport in distal airways would enable further comparison to SPECT-CT data, as we could assess if our model captures how deep particles penetrate into the lung, rather than only regional deposition given here (\fref{fig:sotonDepo}).

Our 0D deep lung model used global resistance and compliance metrics which were based on empirical correlations from experimental measurements \citep{oakes2018airflow, dangelo1989respiratory}.
The resistance and compliance per outlet was then based on the outlet area and the lung's volume fraction (relative to combined volume).
\citet{oakes2018airflow} used this procedure as a initial condition, and then determined the true resistance (accounting for flow resistance in the 3D domain) by iteratively computing the pressure drop between the trachea and each outlet and adjusting the resistance until convergence.
This would require many simulations for each patient to determine the exact distribution of flow and particles delivered to each individual outlet in the domain.
As we are only interested in regional deposition (right and left lung) at this stage to validate our imaging and model predictions, this was not necessary.
Therefore, we have not accounted for 3D domain flow resistance as it has no effect on the distribution of particles between right and left lung, which is dominated by the lobe volume fraction, $\alpha_{(L)}$, and airway morphology.
In future studies where exact local deposition is required, the iterative simulation approach of \citet{oakes2018airflow} may be improved by combining full 3D-0D simulations with a low-fidelity surrogate model to decrease the number of time-consuming simulations \citep{perdikaris2016model}.

\section{Conclusion}
We have developed a rapid image-processing and mathematical modelling pipeline to enable patient-specific predictions of drug deposition with data as sparse as a single chest X-ray image.
This approach segmented lungs and airways from volumetric CT data with high accuracy using convolutional neural networks \citep{ronneberger2015unet, paszke2016enet, hofmanninger2020automatic}.
We also developed an approach to extract lung and airway geometries from chest X-rays using a statistical shape and appearance model that iteratively adjusted shape parameters based on the outline and gray-value distribution of an unseen X-ray image.
This approach was shown to reconstruct patient respiratory morphologies in good agreement with the ground truth data, with the exception of a few outliers.
Deposition in airways from both automated approaches agreed well with deposition in ground truth airways obtained by semi-automatic segmentation.
The SSAM reconstruction failed to reproduce local deposition patterns present in the ground truth, which we expect is due to morphological differences and the reduced number of principal components creating a smoother airway.
Finally, the imaging and modelling framework was compared to experimental \textit{in vivo} measurements \citep{conway2012controlled, fleming2015controlled}.
The quality of images of the upper airway (mouth-throat) was mixed and required cleanup \citep{conway2012controlled}, which introduced uncertainty. 
Despite this uncertainty in the mouth-throat geometry, we achieved good agreement with experimental regional deposition.
To enable future integration of physical models into healthcare settings and e-health frameworks, the ability to generate patient-specific respiratory systems and deposition predictions automatically from sparse data is crucial.
Predicted deposition information could be used as part of the clinical development of new inhaled therapeutics, and reduce the need for expensive and irradiating \textit{in vivo} deposition imaging.
Our developed and experimentally validated framework is an essential step towards clinical implementation of patient-specific modelling, which will allow for automated and reliable predictions of patient therapeutic response to an aerosol drug.

\section*{Acknowledgements}
Simulations reported in this study were performed on Oracle cloud computing platform, funded by Open Clouds Research Environments (OCRE) `Cloud Funding for Research'.
JW was funded by a 2019 PhD Scholarship from the Carnegie-Trust for the Universities of Scotland.
The \textit{in vivo} deposition data used in this study was obtained from a project sponsored by Air Liquide.
The authors thank Daniel Bustamante for his work in performing LUNA16 airway segmentations.

\begin{appendices}

\end{appendices}

\bibliography{main_refs}

\end{document}


\doublespacing

\author[1]{Josh Williams}
\author[1]{Haavard Ahlqvist}
\author[1]{Alexander Cunningham}
\author[2]{Andrew Kirby}
\author[3]{Ira Katz}
\author[4,5]{John Fleming}
\author[4,6]{Joy Conway}
\author[7]{Steve Cunningham}
\author[1*]{Ali Ozel}
\author[1*]{Uwe Wolfram}

\affil[1]{School of Engineering and Physical Sciences, Heriot-Watt University, Edinburgh, UK}
\affil[2]{Royal Hospital for Sick Children, NHS Lothian, Edinburgh, UK}
\affil[3]{Consultant, Meudon, France}
\affil[4]{National Institute of Health Research Biomedical Research Centre in Respiratory Disease, Southampton, UK}
\affil[5]{Department of Medical Physics and Bioengineering, University Hospital Southampton NHS Foundation Trust, Southampton, UK}
\affil[6]{Respiratory Sciences, Centre for Health and Life Sciences, Brunel University, London, UK}
\affil[7]{Centre for Inflammation Research, University of Edinburgh, Edinburgh, UK}
\maketitle
\thanks{* UW and AO share last authorship. \\
Address correspondence to u.wolfram@hw.ac.uk or a.ozel@hw.ac.uk}

\newpage
\section{Manual cleanup of upper airways}

\begin{figure}
	\centering
	\includegraphics{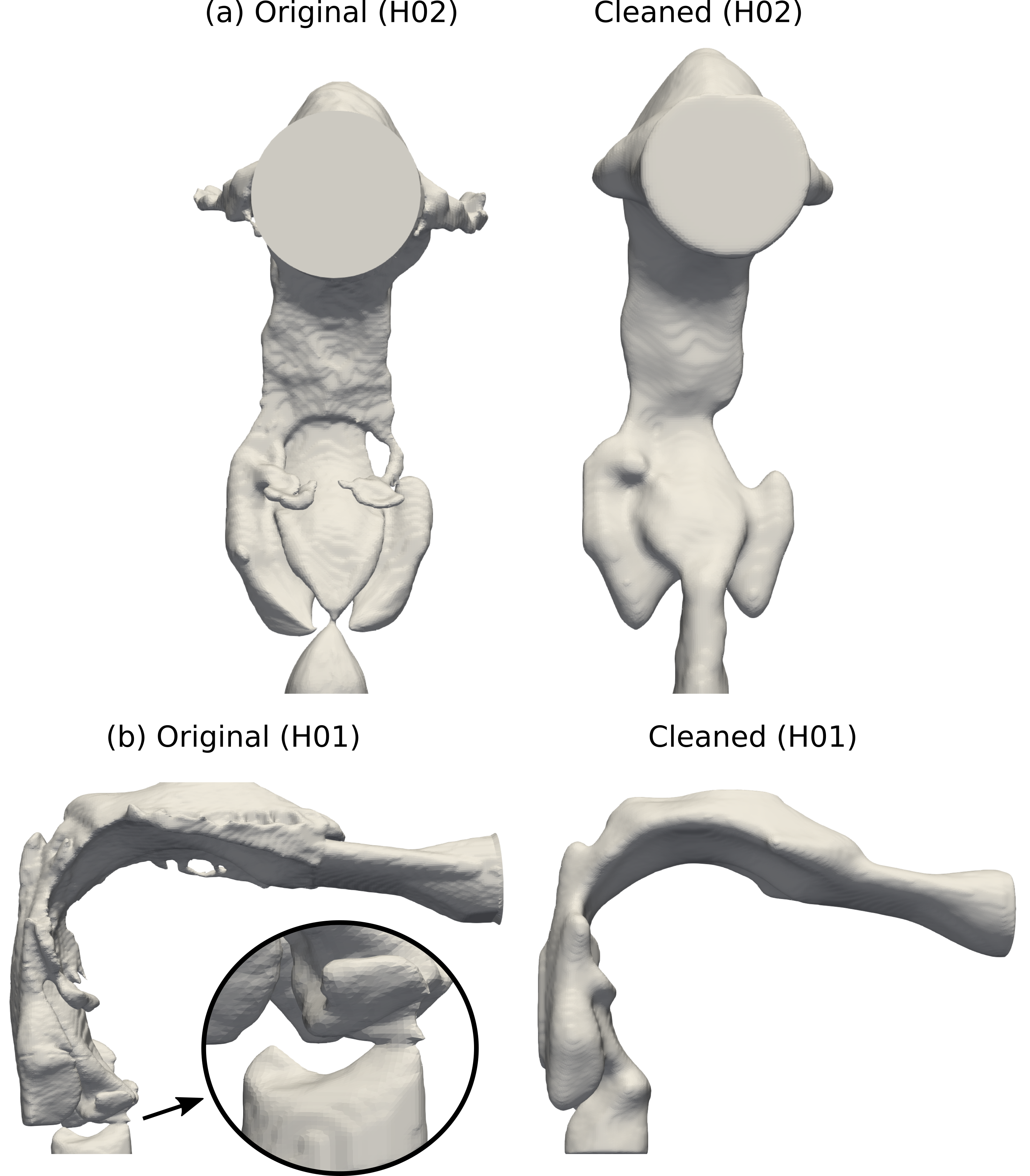}
	\caption[Visualisation of upper airway surface mesh before and after manual cleanup]{Visualisation of upper airway surface mesh before and after manual cleanup for two cases \citep{conway2012controlled}.
	The upper row shows the most constricted case (H02).
	The lower row shows a less extreme, but still heavily constricted case with a zoomed in view.}
	\label{fig:sotonCleanup}
\end{figure}

To mitigate the issues in image acquisition discussed by \citet{conway2012controlled}, we had to manually clean the throat to allow passage for the air and particles to travel through which would be close to expected \textit{in vivo} conditions during inhalation.
We extracted a baseline upper airway segmentation using a semi-automatic region-growing approach implemented in Python. 
The baseline segmentation was imported into 3D Slicer, where the glottal cross-sectional area was filled in manually with a paintbrush tool to better agree with upper airway morphologies in literature \citep{feng2018silico, zhao2020glottis, scheinherr2015realistic}.
The segmentation was then smoothed to remove artificial bumps introduced by the paintbrush tool.
Visual representations of two corrected airways are shown in \fref{fig:sotonCleanup}.

\section{CNN architecture analysis}

\begin{figure}
	\centering
	\includegraphics{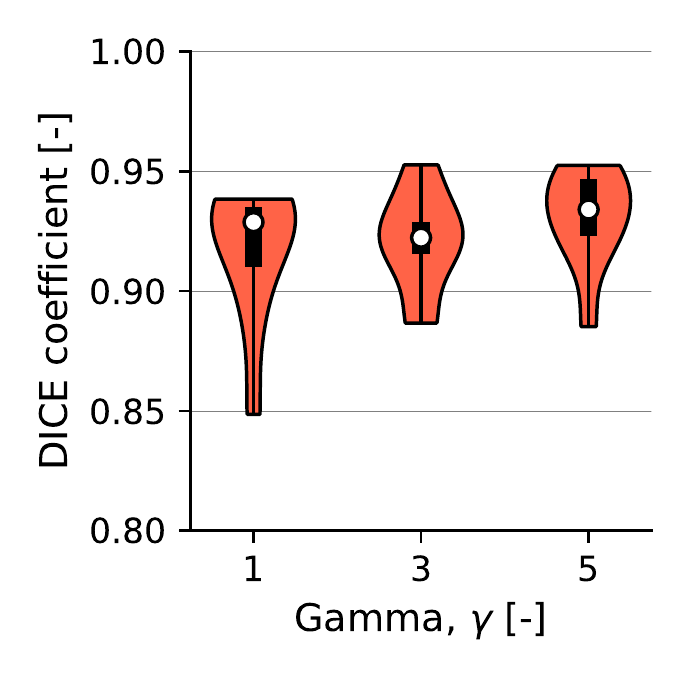}
	\caption[Hyperparameter optimisation for CNN loss function]{Sensitivity analysis of focal loss parameter $\gamma$ in Equation 11 to optimise CNN DICE coefficient. 
	CNNs were trained with no augmentations and results were computed from 10 scans in validation set.}
	\label{fig:unetGamma}
\end{figure}

To find the optimal loss function $\gamma$ parameter in Equation 11, we trained three U-Net CNNs with $\gamma=\{1,3,5\}$ (\fref{fig:unetGamma}).
By comparing the DICE coefficient over the 10 cases in the validation set, we observed $\gamma=1$ to produce the lowest minimum DICE coefficient (0.85). 
When $\gamma=3$ and $\gamma=5$, the minimum and maximum DICE coefficients were approximately the same (0.855 minumum and 0.95 maximum). 
Throughout the remainder of the study, we chose $\gamma=5$ in our loss function due to the slightly higher lower quartile, median and upper quartile (all approximately 1\% larger for $\gamma=5$).

\begin{figure}
	\centering
	\begin{tabular}{cc}
		(a) & (b) \\
		\includegraphics{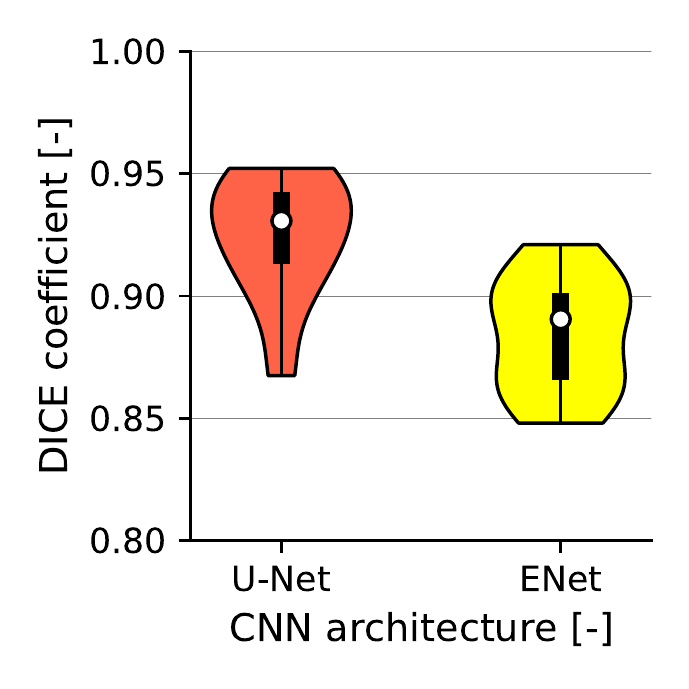} &
		\includegraphics{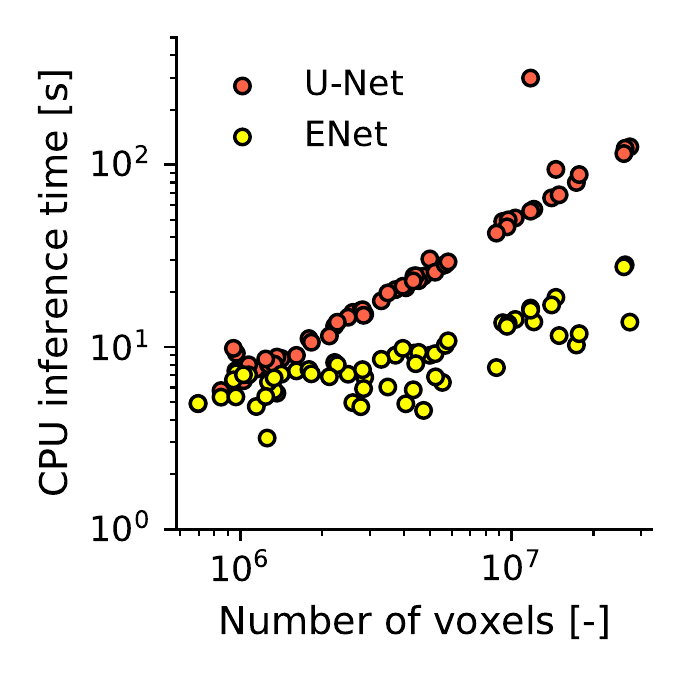}
	\end{tabular}
	\caption[CNN architecture comparison]{Comparison of accuracy and speed for U-Net and ENet architectures. 
	Panels show (a) DICE coefficient from 10 scans in validation set, and (b) inference time relative to size of image, computed on combined training and validation set.}
	\label{fig:cnnArch}
\end{figure}

As the U-Net has a large memory consumption, we also trained an ENet CNN architecture as a less computationally expensive alternative (\fref{fig:cnnArch}a).
Both architectures were trained using $\gamma=5$ and using 25 augmentations (with rotation $\pm 15 ^\circ$). 
There was a significant difference in DICE coefficient ($p < 0.05$), as the U-Net median and upper-quartile was 4\% larger than the ENet. 
The U-Net lower-quartile was 4.7\% larger than the E-Net.
We also compared the time-taken for one forward-pass through the networks (inference time), for all images in the training and validation dataset (\fref{fig:cnnArch}b).
For the smallest images in the dataset (number of voxels below the 25th percentile), the mean inference time was $5.89 \jwunit{s}$ and $7.81 \jwunit{s}$ for the ENet and U-Net, respectively (1.3 times speed-up with ENet).
In contrast, for the largest images in the dataset (number of voxels larger than the 75th percentile), the mean inference time was $15.9 \jwunit{s}$ and $91.28 \jwunit{s}$ for the ENet and U-Net, respectively, which is a speed-up of 5.7 times when using the ENet.
As speed is not a significant driver of our developments at the present stage, we use the U-Net as our default segmentation tool for this study due to its improved DICE coefficient.
However, due to memory overheads required for inference with the U-Net, we used the ENet to segment airways from the high resolution CT of the Southampton/Air Liquide dataset.
Additionally, the ENet DICE coefficient is reasonable and may be preferable for implementation in a clinical workflow, where speed is a key factor.

\section{SSAM morphology assessment}
\begin{figure}
    \centering
    \begin{tabular}{c c}
        (a) One DRR & (b) Two DRRs  \\
        \includegraphics{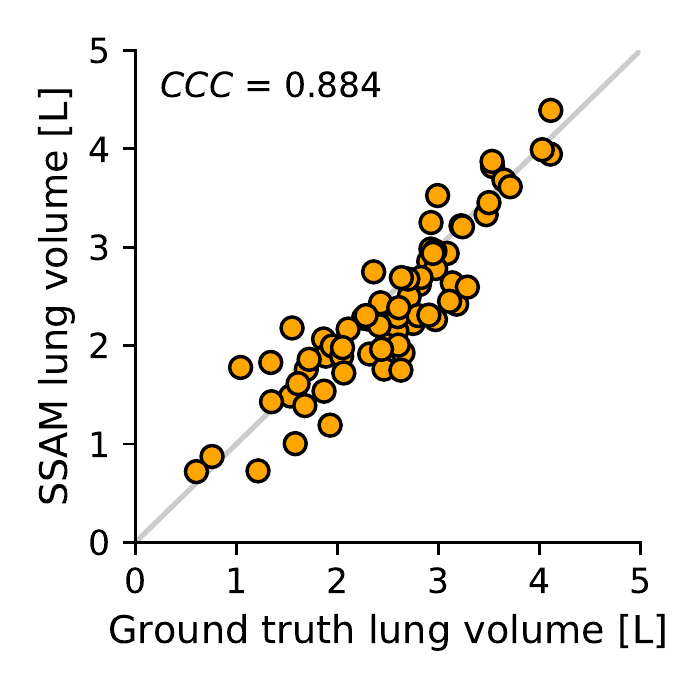} & 
        \includegraphics{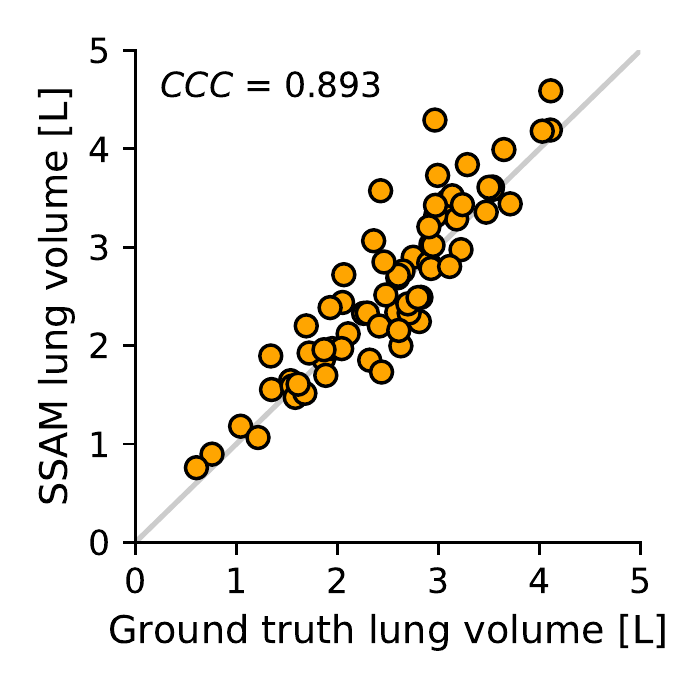}
    \end{tabular}
    \caption{Comparison of lung volume predicted by our SSAM compared to the ground truth segmentations. We show the influence of including an additional projection on lung volume error. Panels show SSAM results with (a) one DRR provided for fitting (anterior-posterior projection), and (b) two DRRs provided (anterior-posterior and lateral projections). $CCC$ is the concordance correlation coefficient.}
    \label{fig:airwayVolLUNA16}
\end{figure}

\begin{figure}
    \centering
    \begin{tabular}{c c}
        (a) One DRR & (b) Two DRRs  \\
        \includegraphics{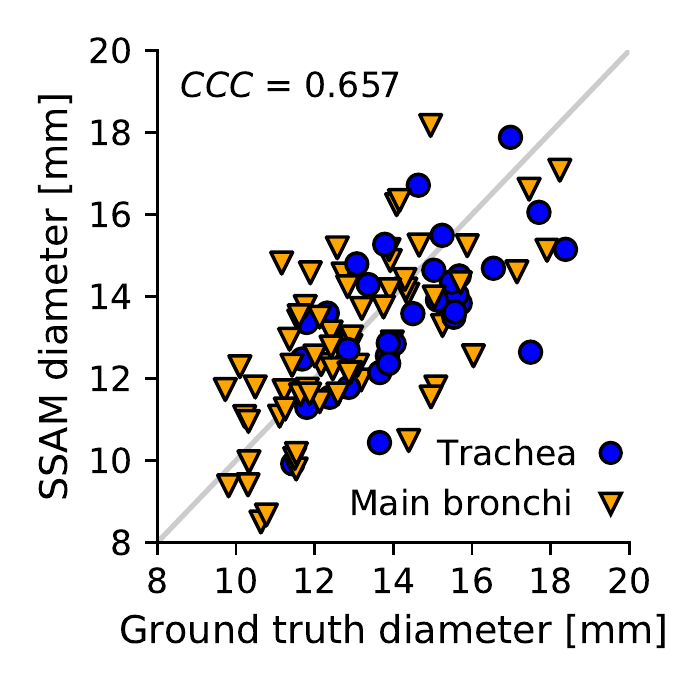} & 
        \includegraphics{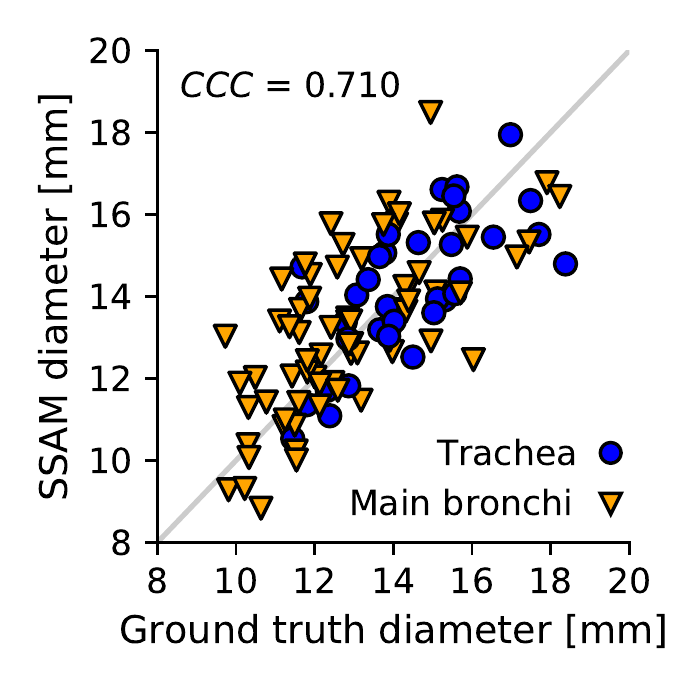}
    \end{tabular}
    \caption{Comparison of airway diameter predicted by our SSAM compared to the ground truth segmentations. We show the influence of including an additional projection on diameter error in the trachea and main bronchi. Panels show SSAM results with (a) one DRR provided for fitting (anterior-posterior projection), and (b) two DRRs provided (anterior-posterior and lateral projections).
    $CCC$ is the concordance correlation coefficient.}
    \label{fig:airwayDiaLUNA16}
\end{figure}

Here we provide the results presented in Figures 3 and 5 as scatter plots to allow for comparison of absolute values of lung space volume and airway diameter. 
We found it important to provide both plots, to show that the absolute values are within a realistic range. This also shows a clear linear correlation between the ground truth measurements and those predicted by the SSAM. 
The lung space volume predicted by the SSAM agreed particularly well (concordance correlation coefficient $CCC \approx 0.89$, \fref{fig:airwayVolLUNA16}).
The diameter predicted by the SSAM (\fref{fig:airwayDiaLUNA16}) showed a less ideal fit than the lung volume, as $CCC=0.657$ and $0.71$. 
The fit was slightly improved with two projections, compared to one projection (5.3\% increase).

\section{Choice of generated airway diameter model}

As the SSAM-reconstructed airways do not have the same level of information as the CT-based airways, best practices for generating distal airways may not be the same as that of airways segmented directly from patient CT data. 
Specifically, on the chest X-ray, only the trachea and main bronchi are visible. 
Even for these `visible' airways, the contrast is poor and the quality of reconstructed diameters was found to be varied in our analysis. 
When using the diameter of the parent branch to assign diameters of the entire airway tree \citep{bordas2015}, it makes sense that erroneous diameters cause some level of error that propagates into the distal airways. Therefore, we compared various models for computing diameter obtained in literature \citep{bordas2015, montesantos2016creation}.

\citet{bordas2015} assigned a logarithmic decay of diameter based on Horsfield order \citep{horsfield1976diameter} defined as
\begin{equation}
\log D_i (H_i) = (H_i - N_H) \log (R_d H) + \log (D_N)
\end{equation}
where $D_i$ is the diameter of branch $i$, $H$ is the Horsfield order, $N_H$ is the Horsfield order of the reference airway in the image-based domain, $D_N$ is the diameter of the reference airway in the image-based domain, $R_d H=1.15$ is a constant representing the logarithmic decrease of diameter with Horsfield order \citep{horsfield1976diameter, tawhai2004, bordas2015}. 
This approach requires declaration of an airway segment in the image-based airways to obtain reference values that are used to assign diameter and Horsfield order to the generated airways. 
\citet{bordas2015} used the parent branch of the image-based airway that is connected to the generated airway, $i$ (referred to here as `parent' model). Alternatively, \citet{montesantos2016creation} calculated airway diameter based on the airway length as $D_i = L_i/3$, with a cutoff $D_{child} \leq 0.95 D_{parent}$ (referred to here as `length' model). Ventilation simulations using the `parent' model was shown to have good agreement with clinical ventilation data in healthy and diseased patients \citep{bordas2015}. Therefore, we used the parent model for CT-based airways.
We used ground truth segmentations with the parent model as a benchmark to evaluate the accuracy of the parent model or length model in on generated airway diameter in SSAM-reconstructed airways.

\begin{figure}
\centering
\begin{tabular}{cc}
    (a) & (b) \\
    \includegraphics[trim={0 0 0 7mm},clip]{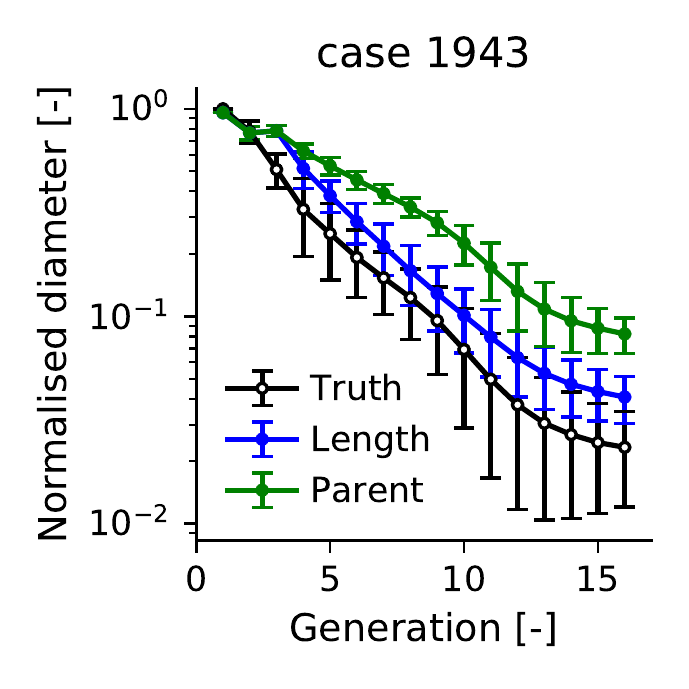} & 
    \includegraphics[trim={0 0 0 7mm},clip]{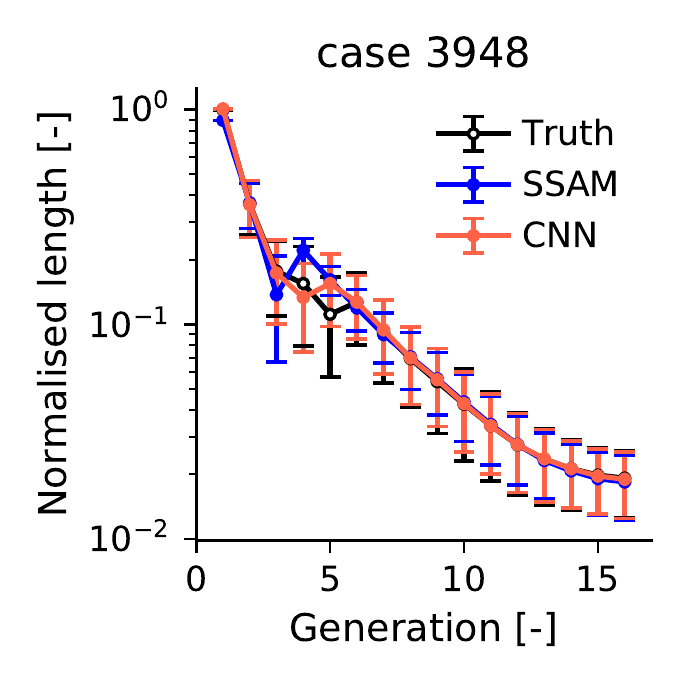}
\end{tabular}
\caption{Comparison of diameter error in full conducting airway tree from a SSAM with varying models to calculate diameter. The `truth' results were generated from ground truth segmentations with the `parent' model for diameter \citep{bordas2015}.}
\label{fig:diameterModel}
\end{figure}

We found the length model to produce better agreement with the ground truth dataset than the parent model proposed by \citet{bordas2015} (\fref{fig:diameterModel}a). 
This is likely due to difficulty inferring the diameter of the central airways from a chest X-ray image. 
As can be seen in \fref{fig:diameterModel}b, the generated airway length showed excellent agreement in the SSAM and ground truth datasets. Based on this, we chose to use the length model of \citet{montesantos2016creation} to assign diameters to the full SSAM airway tree. 

\newpage
\bibliography{main_refs}